\documentclass{article}[12pt]
\usepackage{amsmath,amssymb}
\newtheorem{assumption}{Assumption}
 
\def\eq{\begin{equation}}
\def\en{\end{equation}}
\def\eqa{\begin{eqnarray}}
\def\ena{\end{eqnarray}}
\def\aeq#1{\begin{align}#1\end{align}}  
\def\expval#1{\boldsymbol\langle \, #1 \,\boldsymbol\rangle}
\def\expvalc#1{\expval{#1}_{c}}
\def\expvallambdac#1{\expval{#1}_{\lambda,c}}
\def\Real{{\mathbf{Re}}}

\def\partialby#1{\frac{\partial\hfill}{\partial#1}}
\def\be{\begin{equation}}
\def\ee{\end{equation}}
\def\bea{\begin{eqnarray}}
\def\eea{\end{eqnarray}}

\usepackage{graphicx} 

\begin{document} 
\begin{titlepage}
\def\thepage{}
\ifx\usingvc\undefined{}\else{
\thispagestyle{fancy}                             
}\fi
\begin{center}

\vskip .5in 
{\Large \bf Curvature formula for the
space of 2-d conformal field theories}


\vskip .5in {\large Daniel Friedan}${}^{1}$ and
{\large Anatoly Konechny}${}^{2}$

\vskip 0.5cm
{${}^{1}$Department of Physics and Astronomy\\
Rutgers, The State University of New Jersey\\
Piscataway, New Jersey 08854-8019 USA\\
and\\
Natural Science Institute, The University of Iceland\\
Reykjavik, Iceland\\[0.5ex]
email address: friedan@physics.rutgers.edu\\[3ex]  
${}^{2}$Department of Mathematics,
Heriot-Watt University\\
Riccarton, Edinburgh, EH14 4AS, UK\\
and\\
Maxwell Institute for Mathematical Sciences, 
Edinburgh, UK}\\[0.5ex]
email address: anatolyk@ma.hw.ac.uk
\end{center}

\vskip .5in

\begin{center}June 8, 2012 \end{center}
\vskip .2in
 \large
\begin{abstract} 
We derive a formula for the curvature tensor of the natural Riemannian
metric on the space of two-dimensional conformal field theories and
also a formula for the curvature tensor of the space of boundary
conformal field theories.
\end{abstract}

\end{titlepage}
\newpage \large
\section{Introduction} \large
\renewcommand{\theequation}{\arabic{section}.\arabic{equation}}
\setcounter{equation}{0} 

Following up on the work of Kutasov\cite{Kutasov:1988xb},
we derive a formula
\eq 
R_{ijkl}
=  {\rm RV}\! \int \frac{d^{2}\eta}{2\pi}\; (-\ln |\eta|)\;
\expvalc{\phi_{i}(1)\phi_{j}(\eta)\phi_{k}(\infty)\phi_{l}(0)}
\label{eq:curvature}
\en
for the curvature tensor of 
the Zamolodchikov metric on the space of two dimensional conformal field 
theories (CFT's),
and a similar formula
\aeq{
R_{abcd} &= {\rm RV} \! \int\limits_{-\infty}^{\infty} d\eta\; (-\ln 
|\eta|)\;
\left [
\expvalc{\psi_{a}(1)\psi_{b}(\eta)\psi_{c}(\infty)\psi_{d}(0)}
\right .
\\
&\qquad\qquad\qquad\qquad\qquad
\left . {}+\expvalc{\psi_{a}(0)\psi_{b}(1-\eta)\psi_{c}(\infty)\psi_{d}(1)}
\right ]
\label{eq:bcurvature}
}
for the space of boundary conformal field theories.

In the first formula, the $\phi_{i}$ are exactly marginal 
fields in the conformal field theory at which the 
curvature tensor is calculated.
In the second formula, the $\psi_{a}$ are exactly marginal boundary
fields of the boundary conformal field theory where the 
curvature tensor is calculated.
In both formulas,
$\expvalc{\cdots}$ is the
connected four-point correlation function. 
The integrands have singularities at $\eta=0,1,\infty$.
The letters `RV' denote a particular prescription for regularizing 
and subtracting the divergences --- a hard-sphere (point-splitting) cutoff 
followed by minimal subtraction of the divergences.
The formulas are derived under mild technical assumptions explained 
in section 2 below.
The main limitation is the exclusion of redundant (total 
derivative) fields.
Generically there is no reason to consider redundant fields.
However, as we explain in appendix~\ref{appendix-Cmij},
redundant fields cannot be avoided in a a neighborhood of a CFT with 
continuous symmetry.
In string theory this phenomenon is known as the string Higgs effect.
Appendix~\ref{appendix-Cmij} explains the underlying two-dimensional 
physics.

We study the local geometry abstractly, in terms of the correlation 
functions of the conformal field theory at which we are calculating 
the curvature.
Our work is motivated by a desire to get better control over the
geometry of spaces of conformal field theories and of string theory
vacua.
The N=2 superconformal theories related to
Calabi-Yau manifolds provide well-studied examples of spaces of CFT's.
These 2-d conformal field theories provide string compactifications.
The geometry of their moduli spaces has been determined from consideration
of the low-energy effective field theory corresponding to the 
low-energy string scattering amplitudes.
For these models,
a formula expressing the curvature in terms of the N=2 CFT data
was derived from the low energy effective action \cite{DKL} (see formulas (3.37)
in that paper). More recently in \cite{N2recent} the curvature was computed explicitly 
for a number of examples with N=2 and N=4 supersymmetry.
Our formula~(\ref{eq:curvature}) is general, not restricted to N=2 
CFT's.
We derive the general curvature 
formulas (\ref{eq:curvature}), (\ref{eq:bcurvature}) directly from 2-d conformal field theory in order to 
avoid assuming the low energy effective action.
We are interested in a general derivation directly from 2-d CFT
partly because string theory
requires restrictions on the values of the conformal central charge $c$,
but mainly because there is no complete proof of 
the correspondence between the low energy effective action and the 
string amplitudes.
Our calculations can be considered
as providing a point of support for that correspondence.

We check the curvature formulas in some of the few known families of conformal
field theories where the curvature can be computed directly.

Formulas~(\ref{eq:curvature}) and~(\ref{eq:bcurvature}) can be 
derived in a variety of ways. 
We derive the bulk curvature formula from 
the 2-d conformal anomaly using a slightly novel analytic 
regularization scheme for conformal perturbation theory.
We derive the boundary curvature formula by directly computing second 
derivatives of the metric using a sharp point-splitting cutoff.
We chose such different 
methods  hoping that the techniques might be useful elsewhere. 
We put particular
emphasis on carefully deriving the particular regularization and 
subtraction prescription 
for the integrals in the curvature formulas.

A speculative motivation for deriving the curvature formula is the 
possibility that it could be used to prove
the claim made in \cite{DFlambdamodel} that the natural metric on the 
space of supersymmetric string vacua satisfies an Einstein equation 
$R_{ij} = \frac14 g_{ij}$.

\section{The space of conformal field theories}
\setcounter{equation}{0}

In this paper, a 2-d conformal field theory is a unitary euclidean quantum field theory on the
complex plane.  The trace of the stress-energy tensor vanishes, 
implying locally conserved conformal currents.
The space of conformal field theories is --- modulo some technical
assumptions --- the set of fixed points of the renormalization group
acting on the space of 2-d unitary quantum field theories.
We are interested in the local geometry of the space of conformal field theories 
in the neighborhood of an arbitrary given CFT, the \emph{reference CFT}.
We suppose that the reference CFT has unbroken global conformal 
invariance\footnote{The supposition of global conformal invariance usually goes unspoken.
It avoids the possibility of a locally conformal field theory which, 
on the 2-d plane, exhibits spontaneously broken conformal invariance.
See \cite{FKgradient,Eliezer} for examples and further discussion.
Global conformal 
invariance and unitarity on the plane together imply unitarity of 
the radial quantization.  The self-adjointness of the dilation 
and rotation operators then implies that the local fields can be 
expanded in scaling fields of definite dimension and spin.}
and a discrete spectrum of conformal dimensions.

A family of CFT's in a neighborhood of the reference CFT
is described by coordinates given by
coupling constants
$\lambda^{i}$.
The $\lambda^{i}$
parametrize perturbations of the action of the reference field 
theory that preserve scale invariance.
The partition function of the perturbed theory is
\eq
Z(\lambda) = Z(0) \expval{ e^{\frac1{2\pi}\int {d^{2}z}
\lambda^{i}\phi_{i}(z)} }
\label{eq:bulk-action-principle}
\en
where $Z(0)$ is the partition function of the reference CFT.
The $\phi_{i}(z)$ are local fields in the reference 
CFT, and $\expval{\cdots}$ is the expectation value
in the reference CFT.
%
%

We make the following technical assumptions
\begin{assumption}
The $\phi_{i}(z)$ are dimension 2 scalar fields in the reference CFT.
\label{assumption:dim2}
\end{assumption}
\begin{assumption}
The $\phi_{i}\,\phi_{j}$ operator product expansions (OPE's) contain no 
dimension 2 scalar fields.
\label{assumption:ope}
\end{assumption}
\begin{assumption}
The $\phi_{i}\,\phi_{j}$ OPE's contain no 
dimension 1, spin 1 currents.
\label{assumption:nochiralcurrents}
\end{assumption}
Assumptions~\ref{assumption:dim2} 
and~\ref{assumption:ope} imply that
the beta functions for the couplings $\lambda^{i}$ vanish 
at least through the second order.
Although assumptions~\ref{assumption:ope} 
and~\ref{assumption:nochiralcurrents}
restrict the $\phi_{i}\,\phi_{j}$ OPE's, 
we emphasize that there are
no other restrictions.
In particular, relevant scalar fields can appear in the OPE's.  Scale
invariance is preserved by minimally subtracting the associated power
divergences.  In more general terms we adjust the couplings for the 
relevant fields so their beta functions are zero.
This is especially simple in
the minimal subtraction scheme.  The zeroes of the beta
functions $\beta^{i'}$ for the relevant couplings $\lambda^{i'}$
are at $\lambda^{i'}=0$.
If we used instead a non-minimal scheme, the zeroes of $\beta^{i'}$
could be at non-vanishing values of $\lambda^{i'}$.  For example
in a scheme in which $\beta^{i'} = \delta_{i'}\lambda^{i'} + 
C^{i'}_{jk}\lambda^{j}\lambda^{k}$ we
should set $\lambda^{i'} = - 
C^{i'}_{jk}\lambda^{j}\lambda^{k}/\delta_{i'} $
to preserve conformal invariance.
The non-zero relevant couplings do not contribute to the beta 
function for the marginal couplings by the usual dimensional analysis 
argument --- any such contributions would have negative dimension 
coefficients.

Assumption~\ref{assumption:dim2} excludes any perturbations by 
total derivative operators.  Such a perturbation only amounts to a 
redefinition of the local fields and a corresponding 
reparametrization of the space of CFT's.  None of the physical 
properties change --- the perturbed CFT is equivalent to the 
unperturbed theory.  These perturbations are called 
\emph{redundant}.  In Lagrangian quantum field theory, 
they arise from perturbations by terms that vanish by the equations 
of motion.
Assumption~\ref{assumption:dim2} in conjunction with unitarity and 
global conformal invariance implies that the $\phi_{i}$ are 
primary fields and therefore cannot be total derivatives.

A dimension 1, spin 1 current is necessarily conserved.
If any such current is present in the reference CFT,
assumption~\ref{assumption:nochiralcurrents} states that none of the 
fields $\phi_{i}$ are charged under the corresponding continuous 
symmetry.
If there were such a charged perturbation, it would break the 
continuous symmetry.   We show in Appendix~\ref{appendix-Cmij} that, 
at first order in the symmetry breaking perturbation, a 
certain linear combination of the $\phi_{i}$ becomes a total 
derivative.
Thus assumptions~\ref{assumption:dim2} and~\ref{assumption:nochiralcurrents}
allow us to disregard systematically the possibility of redundant 
perturbations.
One could relax our assumptions to allow for redundant perturbations 
at the cost of technical complication.

We are studying the curvature tensor on a smooth family  of CFT's.
The beta function will vanish identically on such a family, but we 
only need to assume that it vanishes through second order at the 
reference CFT.
This is enough to describe the curvature tensor at a generic point
of the moduli space of CFT's (the space of \emph{all} equivalence 
classes of CFT's).
At generic points the moduli space is  smooth.

Singularities in the moduli space can take various forms.
There are singular points where a number of smooth 
families of CFTs intersect.
Our curvature formula applies to each of the intersecting families.
There are singular points in the moduli space which are CFT's with 
discrete symmetries, under which the perturbations $\phi_{i}$ transform 
nontrivially.  
The discrete symmetries act as equivalence transformations on the smooth family of perturbed theories.
The moduli space of CFT's
is the quotient orbifold.  We are calculating the curvature tensor 
on the smooth family before the discrete quotient is taken.
Another class of singular points in the moduli space arises from CFT's 
with continuous symmetries where some of the perturbations are 
charged.  Again, the symmetries act as equivalence transformations on 
the smooth family of perturbations.
Handling this case would require including redundant operators.

\section{The metric and the curvature tensor}

The natural riemannian metric  $g_{ij}(\lambda)$ on the family of 
CFT's
is extracted from the two-point correlation function
in the 
perturbed CFT,
\eq\label{Zam_metric}
\expval{\phi_{i}(z)\,\phi_{j}(w)}_{\lambda} = g_{ij} (\lambda)\,|z-w|^{-4}
\,.
\en
Scale invariance dictates the form of the two-point function.  The 
coefficient $g_{ij}(0)$ is the riemannian metric at the reference CFT.

To calculate the curvature tensor, we need the first and second 
derivatives of the metric at the reference CFT.  These are calculated in the conformal 
perturbation series, which is
 the expansion of
the partition function and the correlation functions in powers of the
coupling constants $\lambda^{i}$.
The conformal perturbation series is encoded in the generating functional
\eq
Z(\lambda) = Z(0) \expval{ e^{\frac1{2\pi}\int {d^{2}z}
\lambda^{i}(z)\phi_{i}(z)} }
\,,
\en
in which the coupling constants $\lambda^{i}$ in 
equation~\ref{eq:bulk-action-principle} for the partition function
have been replaced by sources
\eq
\lambda^{i}(z) = \lambda^{i} + \delta \lambda^{i}(z)
\en
where the $\delta \lambda^{i}(z)$ have compact support in $z$.
The perturbation series is written
\eq
\ln Z(\lambda) = \ln Z(0) + \sum_{N=1}^{\infty}\ln Z_{(N)}
\en
\eq
\ln Z_{(N)}  = \frac1{N!} 
\int \frac{d^{2}z_{1}}{2\pi}\cdots \frac{d^{2}z_{N}}{2\pi}\;
\lambda^{i_{1}}(z_{1})\cdots \lambda^{i_{N}}(z_{N}) \;
\expvalc{\phi_{i_{1}}(z_{1})\cdots \phi_{i_{N}}(z_{N})}
\en
where the $\expvalc{\cdots}$ are the connected correlation functions in 
the reference CFT.

The connected correlation functions are distributions in the
coordinates $z_{\alpha}$ (so that they can be integrated against the 
sources).
Their singularities are on the diagonals, where some of the 
$z_{\alpha}$ coincide.
Considered as functions of the coordinates $z_{\alpha}$ at 
non-coincident points,
the correlation functions are unambiguously defined.
The integrals of these functions can be singular at coincident 
points,
so renormalization is required to define the correlation functions as 
distributions.
The integrals must be cut off in some 
fashion, then counterterms added to the action so that each
term of the perturbation series goes to a finite limit when the cutoff is removed.
Different renormalization schemes are related by reparametrization of 
the $\lambda^{i}$.  That is, different schemes 
produce different coordinate systems
on the space of conformal field theories.

The expression for the curvature tensor in terms of the derivatives 
of the metric is especially simple in coordinates 
where the first derivatives of the metric vanish:
\eq
R_{ijkl} = 
\frac12 
(\partial_{k}\partial_{j}g_{li}-\partial_{k}\partial_{i}g_{jl}
-\partial_{l}\partial_{j}g_{ki}+\partial_{l}\partial_{i}g_{jk})\,.
\label{eq:curvaturetensor}
\en
Kutasov\cite{Kutasov:1988xb} pointed out that
there is an especially simple renormalization scheme that gives 
such coordinates: the hard-sphere cutoff with minimal subtraction.
The integrals of correlation functions are cut off by restricting
them to the region $|z_{\alpha}-z_{\beta}|>\epsilon$, $\alpha\ne\beta$.
Minimal counterterms depending on $\epsilon$ are added to the action
to cancel the divergences
so that the limit $\epsilon\rightarrow 0$ becomes finite.
The first derivatives of the metric are
\eq
\partial_{k}g_{ij} = \int \frac{d^{2}z}{2\pi} \; 
\expvalc{\phi_{i}(1) \,\phi_{j}(0)\, \phi_{k}(z)}
\,.
\en
The three-point function vanishes identically at non-coincident points,
by assumption~\ref{assumption:ope} (the vanishing of the OPE 
coefficients).
Minimal subtraction means that no finite counterterms are added to the 
action,
so the three-point function vanishes as a distribution,
so the first derivatives of the metric vanish.
As Kutasov remarked, the second derivatives of the metric are
\eq
\partial_{l}\partial_{j}g_{ik} = \int 
\frac{d^{2}z_{1}}{2\pi}\frac{d^{2}z_{2}}{2\pi} \; 
\expvalc{\phi_{i}(1) \,\phi_{k}(0)\, \phi_{l}(z_{1}) \phi_{j}(z_{2})}
\en
so the curvature tensor is given by a sum of 
double integrals of four-point functions.
We take the calculation one step further.
Conformal invariance implies that the four-point function depends, at 
non-coincident points, only 
on one argument, the cross-ratio
\eq
\eta= (1,z_{1};z_{2},0)= \frac{(1-z_{2})z_{1}}{z_{1}-z_{2}}
\,,
\en
so we can perform one of the integrals explicitly, reducing the curvature 
formula to a single integral of the four-point function.
The calculation is complicated by the need for regularization.


\section{The conformal anomaly}
\setcounter{equation}{0}

We find it convenient to calculate the curvature tensor
by extracting the metric
from the integrated conformal anomaly
\eq
\mu\partialby\mu \ln Z(\lambda) = \int\! d^{2}z\, \langle \Theta(z) \rangle \,.
\en
Here $\mu$ is the 2-d scale and $\Theta(z)$ is the trace of the stress-energy tensor.
As a local field, $\Theta(z)$ can be expanded in a basis of 
scaling fields of real dimensions $\ge 0$ and integer spins.
$\Theta(z)$ has canonical dimension $2$ and spin $0$, and the sources 
$\lambda^{i}(z)$ are dimensionless, so the fields that
contribute to $\Theta(z)$ have dimensions 0, 1, and 2.
The only scaling field of dimension 0 is the identity.
Thus the general form of the expectation value of $\Theta(z)$ is \cite{FKgradient,Osborn}
\aeq{
2\pi \expvallambdac{\Theta(z)}
&= \beta^{I}(\lambda)  \expvallambdac{\phi_{I}(z)}
+ C_{i}^{m}(\lambda) \bar \partial \lambda^{i} \expvallambdac{J_{m}(z)}
+ C_{i}^{\bar m}(\lambda) \partial \lambda^{i} \expvallambdac{\bar J_{\bar 
m}(\bar z)}
\label{eq:exptheta}
\\
&\qquad\qquad
{}
+ \partial_{\mu}\left [w_{i}(\lambda) 
\partial^{\mu}\lambda^{i}\right ]
-\frac1{8} g_{ij}(\lambda)\partial_{\mu}\lambda^{i}\partial^{\mu}\lambda^{j}
\nonumber
}
where 
the $\phi_{I}$ are the dimension $\le 2$, spin 
$0$ fields in the reference CFT
and the $J_{m}(z)$, $\bar J_{\bar m}(\bar z)$ are the dimension 
$1$, spin $1$ (chiral) currents in the reference CFT.
The coefficients  on the right hand side are local functionals of the sources,
of appropriate dimension and spin.
The last two terms on the right hand side are proportional to the identity field.
We will check later the appearance of the metric $g_{ij}$ in the 
last term, and its coefficient.
The last four terms on the right hand side comprise the \emph{conformal anomaly} (in 
a flat 2-d geometry).

The beta functions $\beta^{I}(\lambda)$ of course vanish identically on a family of CFT's
so the first term on the right hand side does not occur.
But our assumptions only require that the $\beta^{I}(\lambda)$ 
vanish through second order.
To calculate the curvature tensor, we will expand 
equation~(\ref{eq:exptheta})
to fourth order in the sources.
The fourth derivative of $\beta^{I}(\lambda)$ will multiply a 
one-point function, which vanishes.
The third derivative of $\beta^{I}(\lambda)$ will be symmetric in the 
three indices, so cannot contribute to the curvature tensor.
So we can ignore the first term on the right hand side 
of~(\ref{eq:exptheta}).
To avoid cluttering the calculations, we will take the third derivatives of $\beta^{I}(\lambda)$
to be zero.  As we have argued, the result for the curvature tensor 
is not affected.

Equation~(\ref{eq:exptheta}) implies
the OPE in the reference CFT of the form
\eq
T(z)\, \phi_{i}(0) \sim \frac1{z^{3}} C^{\bar m}_{i}(0) \bar J_{\bar 
m}(0) + \cdots
\en
where $T(z)$ is the usual holomorphic component of the stress-energy 
tensor.  Such a term is forbidden by global conformal invariance and 
unitarity.
Therefore
$C^{\bar m}_{i}(0) = 0$, and similarly $C^{m}_{i}(0) = 0$.
In Appendix~\ref{appendix-Cmij} 
we show that the first derivatives
$C^{m}_{ij}= \partial_{i}C^{m}_{j}(0)$ and
$C^{\bar m}_{ij}= \partial_{i}C^{\bar m}_{j}(0)$
appear as operator product coefficients
\eq\label{OPEJ}
\phi_{i}(z)\,\phi_{j}(0) \sim |z|^{-4} 
\left [ z \, C^{m}_{ij} J_{m}(0) + \bar z\, C^{\bar m}_{ij} \bar J_{\bar m}(0) \right ]
\en
and thus vanish by assumption~\ref{assumption:nochiralcurrents}.
This is enough to show that the second and third terms on the right 
hand side of~(\ref{eq:exptheta}) do not contribute to the curvature 
calculation.
The fourth derivatives of $C^{m}_{i}(\lambda)$
multiply $\expval{J_{m}}_{0,c}$ which vanishes.
The third derivatives multiply two-point functions
$\expval{J_{m}\phi_{k}}_{0,c}$ which vanish.
Finally, the second derivatives of $C^{m}_{i}(\lambda)$
multiply three-point functions
$\expval{J_{m}\phi_{j}\phi_{k}}_{0,c}$ which vanish by 
assumption~\ref{assumption:nochiralcurrents}.
The same holds for $C^{\bar m}_{i}(\lambda)$.

The fourth term in~(\ref{eq:exptheta}) is a total derivative
so we can write
\eq 
\label{anomaly}
\mu\partialby\mu \ln Z(\lambda)
= 
\int d^{2}z\; \expvallambdac{\Theta(z)}
= 
-\int \frac{d^{2}z}{2\pi} 
\frac18 \;g_{ij}(\lambda)
\,\partial_{\mu}\lambda^{i} \partial^{\mu}\lambda^{j} 
+ \cdots
\en
where the omitted terms make no contribution to the curvature tensor.

The tensor $g_{ij}(\lambda)$ in (\ref{anomaly}) is the Zamolodchikov metric 
(\ref{Zam_metric}).
This is derived
by noting that, with the hard-sphere regularization, the divergent part of
\eq
\frac12  \int \frac{d^{2}z_{1}}{2\pi} \; \frac{d^{2}z_{2}}{2\pi} \; 
\theta(|z_{1}-z_{2}|-\epsilon)
\;\lambda^{i}(z_{1}) \lambda^{j}(z_{2}) \expvalc{\phi_{i}(z_{1}) \,\phi_{j}(z_{2})}
\en
is cancelled by the counterterms
\eq
\Delta S = \int \frac{d^{2}z}{2\pi}
\;
\left [
\epsilon^{-2} \frac14 g_{ij}\lambda^{i}\lambda^{j}
+\ln (\mu\epsilon) \frac18 \;g_{ij}
\,\partial_{\mu}\lambda^{i} \partial^{\mu}\lambda^{j} 
\right ]
\en
so
\eq
\mu\partialby\mu \ln Z(\lambda)
= 
-\int \frac{d^{2}z}{2\pi} 
\frac18 \;g_{ij}
\,\partial_{\mu}\lambda^{i} \partial^{\mu}\lambda^{j} 
\en
to second order in the sources $\lambda^{i}(z)$.
This local calculation works as well in any nearby conformal field theory,
so the integrated anomaly must be as in equation \ref{anomaly}.
The equation does not depend on the renormalization scheme because no finite counterterms can affect 
it.

\section{The curvature tensor}
\setcounter{equation}{0}

The second derivatives of the metric are now found by expanding the 
anomaly to fourth order in the $\lambda^{i}$,
\eq
\mu\partialby\mu\ln Z_{(4)}
= 
-\int \frac{d^{2}z}{2\pi} 
\frac1{16} \partial_{k}\partial_{l} \;g_{ij}
\,\lambda^{k}\lambda^{l}\partial_{\mu}\lambda^{i} \partial^{\mu}\lambda^{j} 
\en
where the fourth order term in the conformal perturbation series is
\eq
\ln Z_{(4)}
=
\frac1{4!} \int \prod_{\alpha=1}^{4} \frac{d^{2}z_{\alpha}}{2\pi}
\; \expvalc{\prod_{\alpha=1}^{4}\lambda^{i}(z_{\alpha}) 
\phi_{i}(z_{\alpha})}
\,.
\en
Changing integration variables to $y_{\alpha}=z_{\alpha} - z$
with $z= \sum_{\alpha}z_{\alpha}$,
then expanding each $\lambda^{i_{\alpha}}(z+y_{\alpha})$ in powers of 
the $y_{\alpha}$, keeping the 
terms containing two derivatives of the sources,
gives
\aeq{
\mu\partialby\mu\ln Z_{(4)} &=
- \int d^{2}z\; 
\lambda^{i_{2}}\lambda^{i_{3}}\partial_{\mu}\lambda^{i_{1}}\partial^{\mu}\lambda^{i_{4}}
\;\frac1{16}
\int 
\prod_{\alpha=1}^{4} 
\frac{d^{2}y_{\alpha}}{2\pi}\; \delta^{2}\left (\frac14\sum y_{\alpha}\right )
\; \left | y_{1}-y_{4}\right |^{2}
\\ &\qquad\qquad\qquad\qquad
\mu\partialby\mu
\expvalc{\phi_{i_{1}}(y_{1})\phi_{i_{2}}(y_{2})\phi_{i_{3}}(y_{3})\phi_{i_{4}}(y_{4})}
}
from which we can read off the second derivatives of the metric
\aeq{
\partial_{i_{2}}\partial_{i_{3}}g_{i_{1}i_{4}}
&=
\int 
\prod_{\alpha=1}^{4} 
\frac{d^{2}y_{\alpha}}{2\pi}\;
2\pi \delta^{2}\left ( \frac14 \sum y_{\alpha}\right )
\; \left | y_{1}-y_{4}\right |^{2}
\\
&\qquad\qquad\qquad\qquad\qquad
\mu\partialby\mu
\expvalc{\phi_{i_{1}}(y_{1})\phi_{i_{2}}(y_{2})\phi_{i_{3}}(y_{3})\phi_{i_{4}}(y_{4})}
\,.
}
Substituting in equation (\ref{eq:curvaturetensor}), we obtain
\aeq{
R_{i_{1}i_{2}i_{3}i_{4}} 
&=
\frac12
\int 
\prod_{\alpha=1}^{4} 
\frac{d^{2}y_{\alpha}}{2\pi}\;
2\pi \delta^{2}\left ( \frac14 \sum y_{\alpha}\right )
\\
&\qquad\qquad\qquad\qquad\qquad 
\left ( \left | y_{1}-y_{4}\right |^{2}- \left | y_{2}-y_{4}\right |^{2}
-\left | y_{1}-y_{3}\right |^{2}+\left | y_{2}-y_{3}\right |^{2}
\right )
\\
&\qquad\qquad\qquad\qquad\qquad
\mu\partialby\mu
\expvalc{\phi_{i_{1}}(y_{1})\phi_{i_{2}}(y_{2})\phi_{i_{3}}(y_{3})\phi_{i_{4}}(y_{4})}
\,.
}
Changing variables from $y_{\alpha}$ to $x_{\alpha}= y_{\alpha}-y_{4}$, 
$\alpha=1,2,3$,
and integrating over $y_{4}$, we obtain
\eq
R_{i_{1}i_{2}i_{3}i_{4}} 
=
\int 
\prod_{\alpha=1}^{3} 
\frac{d^{2}x_{\alpha}}{2\pi}\;
\left [ \left (  x_{1}-x_{2}\right ) \cdot x_{3} \right ]
\mu\partialby\mu
\expvalc{\phi_{i_{1}}(x_{1})\phi_{i_{2}}(x_{2})\phi_{i_{3}}(x_{3})\phi_{i_{4}}(0)}
\label{eq:curv1}
\en
where we write
\eq
u\cdot v = \frac12 \left (\bar u v + u \bar v \right ) = \Real (\bar 
u v)\,.
\en
The scale derivative of the four-point correlation function is
the fourth variation of the integrated anomaly with respect to the 
sources.
By the arguments of the previous section,
the scale derivative vanishes away from 
the diagonal $x_{1}=x_{2}=x_{3}=0$.
Thus the region of integration can be restricted to any region that
includes the diagonal.


To construct the renormalized four-point 
function we use a version of analytic regularization.
We define the regulated N-point function
\eq
G_{s}({ \bf z}) = 
\mu^{Ns}K_{s}({ \bf z})\langle \phi_{i_{1}} (z_{1})\dots 
\phi_{i_{n}}(z_{n}) \rangle_{c}\, .
\en
where
 \eq
K_{s}({\bf z}) =  \prod_{\alpha<\beta} 
|z_{\alpha}-z_{\beta}|^{\frac{2s}{N-1}}
\,.
\en
The crucial point of this definition is that the regulated fields $\phi_{i}$ have 
scaling dimension $2-s$, as in dimensional regularization of 
lagrangian quantum field theory.  Then $\mu^{s} \phi_{i}$ has dimension 2
so we still have the canonical scaling relation
\eq
\left (
\mu\partialby\mu
-
\sum_{\alpha} z_{\alpha} \cdot \partialby{z_{\alpha}}
-2N
\right ) 
G_{s}({\bf z})
=
0\,.
\label{eq:canonicalscaling}
\en

For $\Real \, s >1$,
the regulated correlation functions $G_{s}({\bf z})$ are nonsingular
distributions in the coordinates $z_{\alpha}$.
They are holomorphic functions of 
the regularization parameter $s$
which analytically continue to meromorphic functions of $s$.
The renormalized correlation 
functions are obtained by subtracting poles at $s=0$ and taking the limit $s\to 0$
\eq
\langle \phi_{i_{1}} (z_{1})\dots 
\phi_{i_{n}}(z_{n}) \rangle_{c}
= \lim_{s\rightarrow 0} \left [ G_{s}({\bf z}) - \Delta G_{s}({\bf 
z}) \right ]
\en
where the counterterm $\Delta G_{s}({\bf z}) $
contains poles at $s=0$ and is independent of $\mu$.
Thus
\eq
\mu\partialby\mu
\langle \phi_{i_{1}} (z_{1})\dots 
\phi_{i_{n}}(z_{n}) \rangle_{c}
=
\lim_{s\rightarrow 0} \mu\partialby\mu G_{s}({\bf z})
\,.
\en
Equation (\ref{eq:curv1}) becomes
\eq
R_{i_{1}i_{2}i_{3}i_{4}} 
=
\lim_{s\rightarrow 0}
\int 
\prod_{\alpha=1}^{3} 
\frac{d^{2}x_{\alpha}}{2\pi}\;
\left [ \left (  x_{1}-x_{2}\right ) \cdot x_{3} \right ]
\mu\partialby\mu
G_{s}(\mathbf{x})
\en
where $G_{s}(\mathbf{x})$ now stands for the regulated four-point function 
\eq
G_{s}(\mathbf{x}) 
= \mu^{4s} K_{s}(\mathbf{x}) \expval{\phi_{i_{1}}(x_{1})\phi_{i_{2}}(x_{2})\phi_{i_{3}}(x_{3})\phi_{i_{4}}(0)}_{c}
\en
\eq
K_{s}(\mathbf{x}) = \left |
x_{1}x_{2}x_{3}(x_{3}-x_{1})(x_{3}-x_{2})(x_{1}-x_{2})
\right |^{\frac{2s}3} \, .
\en
By (\ref{eq:canonicalscaling}),
\eq \label{curv_eq}
R_{i_{1}i_{2}i_{3}i_{4}} 
=
\lim_{s\rightarrow 0}
\int 
\prod_{\alpha=1}^{3} 
\frac{d^{2}x_{\alpha}}{2\pi}\;
\sum_{\alpha}\partialby{x_{\alpha}}\cdot
\big ( x_{\alpha}\,
\left [ \left (  x_{1}-x_{2}\right ) \cdot x_{3} \right ]
G_{s}(\mathbf{x})
\big )
\,.
\en

Since the integral in (\ref{curv_eq}) vanishes off the diagonal, 
we can introduce ---  without affecting the result ---
a factor $B(\mathbf{x})$ in the integrand
that equals 1 in a neighborhood of $(0,0,0,0)$ and 
drops off sufficiently fast at infinity. We pick 
\eq\label{B}
B(\mathbf{x}) = e^{-\epsilon^{2} \lVert \mathbf{x} \rVert^{2}}
\en
with 
\eq
\lVert \mathbf{x} \rVert^{2} = 
|x_{3}|^{2}+|x_{1}|^{2} + |x_{2}-x_{3}|^{2}\, .
\en
Integrating by parts in (\ref{curv_eq}) we obtain 
\aeq{
R_{i_{1}i_{2}i_{3}i_{4}} 
&=
\lim_{s\rightarrow 0}\;
\int 
\prod_{\alpha=1}^{3} 
\frac{d^{2}x_{\alpha}}{2\pi}\;
B(\mathbf{x})
\sum_{\alpha}  \partial_{\alpha}
\cdot
\big (x_{\alpha}  \left [\left (  x_{1}-x_{2}\right ) \cdot x_{3}\right ] 
G_{s}(\mathbf{x})\big )
\\
&=
\lim_{s\rightarrow 0}\;
\int 
\prod_{\alpha=1}^{3} 
\frac{d^{2}x_{\alpha}}{2\pi}\;
(-DB)(\mathbf{x})
\left [\left (  x_{1}-x_{2}\right ) \cdot x_{3}\right ] 
G_{s}(\mathbf{x})
}
where
\eq
DB(\mathbf{x}) = \sum_{\alpha}   x_{\alpha} \cdot \partial_{\alpha} B(\mathbf{x})\, .
\en

The regulated correlation function depends only on the correlation 
function at non-coincident points, which is invariant under global 
conformal transformations,
so we can rewrite the last formula as 
\begin{multline}
R_{i_{1}i_{2}i_{3}i_{4}} 
=
\lim_{s\rightarrow 0}\;
\int 
\prod_{\alpha=1}^{3} 
\frac{d^{2}x_{\alpha}}{2\pi}\;
(-DB)(\mathbf{x})
\\
\left [\left (  x_{1}-x_{2}\right ) \cdot x_{3}\right ] 
\left | x_{1}(x_{3}-x_{2}) \right |^{-4}
K_{s}(\mathbf{x})
G(1,\eta,\infty,0)
\end{multline}
where 
\eq
G(1,\eta,\infty,0)=\langle  
\phi_{i_{1}}(1)\phi_{i_{2}}(\eta)\phi_{i_{3}}(\infty)\phi_{i_{4}}(0)  
\rangle_{c}
\en
and 
\eq
\eta = \frac{(x_{3}-x_{1})x_{2}}{(x_{3}-x_{2})x_{1}}
\en
is the cross-ratio. 
We have dropped the factor $\mu^{4s}$ from the regulated four-point 
function because the limit $s\rightarrow 0$ is finite.

We now have the curvature formula as a single integral
\eq
R_{i_{1}i_{2}i_{3}i_{4}} 
=
\lim_{s\rightarrow 0}\;
\int\frac{d^{2}\eta}{2\pi}\;
f_{s}(\eta) G(1,\eta,\infty,0)
\en
with
\begin{multline}
\label{fs1}
f_{s}(\eta) =
\int 
\prod_{\alpha=1}^{3} 
\frac{d^{2}x_{\alpha}}{2\pi}\;
2\pi \delta^{2}\left (
\eta - \frac{(x_{3}-x_{1})x_{2}}{(x_{3}-x_{2})x_{1}}
\right )
(-DB)(\mathbf{x})
\\
\left [\left (  x_{1}-x_{2}\right ) \cdot x_{3}\right ] 
\left | x_{1}(x_{3}-x_{2}) \right |^{-4}
K_{s}(\mathbf{x})\, .
\end{multline}
Changing the variables of integration to $u_{1}=x_{1}/x_{3}$, $u_{2}=x_{2}/x_{3}$ and $x_{3}$ and using (\ref{B}) we can perform 
the integration over $x_{3}$ in (\ref{fs1}) to get
\begin{multline}
f_{s}(\eta)=
\frac{1}{2\pi}\int\!\!
d^{2}u_{1}\;
d^{2}u_{2}\;
\delta^{2}\left (
\eta - \frac{u_{1}^{-1}-1}{u_{2}^{-1}-1}
\right )
\\
\Real (  u_{1}-u_{2} ) 
\left | u_{1}(1-u_{2}) \right |^{-4}
K_{s}(\mathbf{u})
\Gamma(2s+1)\epsilon^{-4s} \lVert \mathbf{u} \rVert^{-4s}
\end{multline}
where 
\eq
\lVert \mathbf{u} \rVert^{2} = 1 + |u_{1}|^{2}+|1-u_{2}|^{2}\, ,
\en
\eq
K_{s}(\mathbf{u}) = \left |
u_{1}u_{2}(1-u_{1})(1-u_{2})(u_{1}-u_{2})
\right |^{\frac{2s}3}\, .
\en
Performing a further change of variables 
\eq
v=\frac{1}{u}-1\, , \qquad w=\left(\frac{1}{u_{2}}-1\right)^{-1}
\en
and using the delta function to integrate out $w$ we obtain 
\eq
f_{s}(\eta)= \Gamma(2s+1)\epsilon^{-4s} 
|\eta(\eta-1)|^{\frac{2s}3} \;\Real\; g_{s}(\eta)\, , 
\en
\eq
\label{eq:gs}
g_{s}(\eta) =(\eta-1)\int
\frac{d^{2}v}{2\pi}\;
\left | v\right |^{-2}
\left [
(1-v)(\eta-v)v^{-1}
\right ]^{-1}
\left |
(1-v)\left (\eta-v\right)v^{-1}
\right |^{-2s}
\lVert \mathbf{u} \rVert^{-4s}\, , 
\en
\eq
\label{eq:u}
\lVert \mathbf{u} \rVert^{2} = |1-v|^{-2}+|1-\eta^{-1}v|^{-2}+1 \, .
\en

The function
$f_{s}(\eta)$ is integrated against the four-point function $G(1,\eta,
\infty, 0)$ which has singularities at $\eta=0,1,\infty$.
Away from those three points, as we will see shortly,
\eq
\lim_{s\rightarrow 0} f_{s}(\eta) = - \ln |\eta|
\,.
\en
Near the singular points $0,1,\infty$, we have to perform the 
integral
with $s>0$ then take the limit $s\rightarrow 0$.
Because the function $f_{s}(\eta)$ is so complicated,
it is not immediately obvious how the regularization works.
We will make the regularizing effect of $f_{s}(\eta)$ explicit by 
analyzing the integral in the immediate neighborhood of the singular 
points.
We will then replace the regularization by $f_{s}(\eta)$ by an 
equivalent but much 
simpler prescription which uses the hard-sphere 
regularization\footnote{We did not use hard-sphere regularization 
from the beginning because, as we will see later for boundary CFT,
the reduction of the curvature formula to a single 
integral using hard-sphere 
regularization is a complicated calculation.}.

To see how
the singularities are regularized we need to know the behaviour of
$f_{s}(\eta)$ near $\eta=0,1,\infty$.  The analysis is somewhat
tedious but straightforward\footnote{It is advantageous to split the
regions of $v$-integration into 3 parts.  Thus for $\eta\to 0 $ we can
take $0\le |v|\le |\eta|/\rho$, $|\eta|/\rho \le |v|\le \rho$, $\rho
\le |v|<\infty$ where $\rho$ is any real number such that
$|\eta|<\rho<1$.  For $\eta$ tending to $\infty$ and 1 same type of
splitting is obtained after first changing the variables as $v\to
v^{-1}$ and $v\to 1-v$ respectively.}.  We find the following
expressions
\be\label{fs_asympt}
f_{s}(\eta) = \left \{
\begin{array}{l@{\qquad}l}
|\eta|^{\frac{2s}3} \;
\left [
-A^{(0)}_{s}(\eta) \ln |\eta| +B^{(0)}_{s}(\eta)
\right ]
&  |\eta|<1
\\[2ex]
|1-\eta|^{\frac{2s}3}  \Real\,\left [ (1-\eta) A^{(1)}_{s}(1-\eta)\right ] 
& |1-\eta|<1 \\[2ex]
|\eta|^{-\frac{2s}3}  \left [ 
-A^{(\infty)}_{s}(\eta^{-1}) 
\ln |\eta| 
+B^{(\infty)}_{s}(\eta^{-1}) \right ]
& 1< |\eta|
\end{array}
\right .
\ee
where $ A^{(0,\infty)}_{s}$ and $ B^{(0,\infty)}_{s}$ are real-valued,
$A^{(1)}_{s}$ is complex valued,
and all five functions are real-analytic in $\eta$ in the appropriate 
domains, for $s>0$.
All are regular in $s$ for $\Real\, s > -\frac12$.  At $s=0$ we have
\be
A^{(0)}_{0}(\eta)  =A^{(\infty)}_{0}(\eta)  =1\,,
\quad
B^{(0)}_{0}(\eta)  = B^{(\infty)}_{0}(\eta)  =0\,,
\quad
A^{(1)}_{0}(1-\eta)  = - \eta^{-1} \ln (1-\eta)
\,.
\ee

Singularities of the four-point function $G(1,\eta, \infty, 0)$ arise
from three sources.  At $\eta=0$, the relevant spin 0 fields give
singularities which go as $|\eta|^{\Delta - 4}$ with $0<\Delta < 2$.
While we excluded chiral spin 1 fields from the OPE's, the non-chiral
spin 1 fields contribute divergences $|\eta|^{\delta-4}\eta$ with
$0<\delta<1$.  Finally, the fields of spin 2 and dimension 2 contribute
singularities that go as $|\eta|^{-4}\eta^{2}$.  Using the expressions
(\ref{fs_asympt}) we find that all of these singularities are
regularized when multiplied by $f_{s}(\eta)$ as long as $\Real\, s>0$.
For the contribution of a relevant field, we find
\eq
\frac1{2\pi}\int_{|\eta|<a} d^{2}\eta \; f_{s}(\eta) \, |\eta|^{\Delta-4}
= 
\frac{-a^{\Delta-2}\ln a}{\Delta-2} 
+ \frac{a^{\Delta-2}}{(\Delta-2)^{2}} 
+ \cdots
\label{eq:counterterm}
\en
where the ommitted terms vanish as $a\rightarrow 0$.
The contributions of spin 1 and spin 2 fields vanish by rotation 
invariance.  We thus see that the regularization by $f_{s}(\eta)$ is 
equivalent to the hard-sphere cutoff plus minimal counterterms of 
the form (\ref{eq:counterterm}) for each relevant scalar field that occurs in the 
$\phi_{i_{2}}(\eta) \,\phi_{i_{4}}(0)$ OPE.
In the absence of relevant scalar fields in the OPE, there are no 
counterterms at $\eta=0$ and we simply get 
the principal value prescription.

The analysis at $\eta = \infty$ is exactly the same.  The hard-sphere 
cutoff is $|\eta|<a^{-1}$.  The integral near $\infty$, over the 
region $|\eta|\ge a^{-1}$, contributes minimal counterterms for each 
relevant scalar in the $\phi_{i_{2}}(\eta)\phi_{i_{3}}(\infty)$ OPE.

Finally, the extra factor $\eta-1$ in the 
asymptotics of $f_{s}(\eta)$ as $\eta\rightarrow 1$ 
means that the integral over
the region $|\eta-1|\le a$ is finite in the limit $a\rightarrow 0$,
so we just have the principal value prescription at $\eta =1$.

We have obtained
\begin{multline}
\label{finalR}
R_{i_{1}i_{2}i_{3}i_{4}} 
= {\rm RV}
\int 
\frac{d^{2}\eta}{2\pi}\;
( - \ln |\eta|)\,
\expvalc{\phi_{i_{1}}(1)\phi_{i_{2}}(\eta)\phi_{i_{3}}(\infty)\phi_{i_{4}}(0)}
\\
=\lim _{a\rightarrow 0}
\bigg [
\int\limits_{\stackrel{a<|\eta|<a^{-1}}{\scriptscriptstyle a< |1-\eta|}}
\frac{d^{2}\eta}{2\pi}\;
( - \ln |\eta|)\,
\expvalc{\phi_{i_{1}}(1)\phi_{i_{2}}(\eta)\phi_{i_{3}}(\infty)\phi_{i_{4}}(0)}
+\Delta R_{i_{1}i_{2}i_{3}i_{4}}(a)
\bigg ]
\end{multline}
where $\Delta R_{i_{1}i_{2}i_{3}i_{4}}(a)$ are the minimal 
counterterms due to relevant scalars, as explained above.

This formula
was obtained using a regularization in which the first
derivatives of the metric vanish.
Our final formula (\ref{finalR})
depends only on the values of the four-point functions at finite
separations,
therefore it transforms covariantly as a 4-tensor.
Therefore (\ref{finalR}) 
is coordinate-independent.

It is slightly nontrivial to check the symmetry properties of $R_{i_{1}i_{2}i_{3}i_{4}} $
given by (\ref{finalR}) and the first Bianchi identity.
Formally they follow directly from invariance of the four-point function 
under the conformal transformations that permute $0,1,\infty$,
but the regularization is not manifestly conformally invariant.
Under our assumptions 1-3, it turns out that 
that the regularization does not spoil
the global conformal symmetries.

\section{Two-dimensional torus example}
To check the curvature formula we look at the moduli space of the two-dimensional torus CFT.
This model can be described in terms of   a free  complex  bosonic field $X(z, \bar z)$ subject to identifications 
\be
X\sim X + 2\pi\, , \quad X \sim X + 2\pi i
\ee
The action is 
\be 
S = \int\frac{d^{2}z}{2\pi i}(\tau \partial X\bar \partial X^{*} - \bar \tau \bar \partial X\partial X^{*})
\ee
where $X^{*}$ is the complex conjugate field and $\tau$ is the coupling constant that specifies the Kahler form on the 
target space two-torus. 
We are considering the family of CFT's parametrized by $\tau$.
For simplicity we hold fixed the target space complex structure. 

The propagator is 
\be \label{XXprop}
\langle X^{*}(z, \bar z)X(0)\rangle = -\frac{1}{{\rm Im}\,\tau}\ln|z|^{2}\, .
\ee
The variation of the Kahler modulus $\tau$ is described by the action variation 
\be 
\delta S = -\int\frac{d^{2}z}{2\pi }(\delta \tau \phi_{\tau} + \delta \bar \tau \phi_{\bar \tau})
\ee
where
\be
\phi_{\tau}=-i\partial X \bar \partial X^{*}\, , \qquad \phi_{\bar \tau}=i\bar \partial X  \partial X^{*} \, .
\ee
The two-point function 
\be
\langle \phi_{\bar \tau}(z, \bar z)\phi_{\tau}(0)\rangle = ({\rm Im}\,\tau)^{-2} |z|^{-4}
\ee
gives the Zamolodchikov metric 
\be
ds^{2}
= g_{\bar \tau \tau} |d\tau|^{2}+g_{\tau \bar \tau} |d\tau|^{2}
=2({\rm Im}\, \tau)^{-2}|d\tau|^{2}\,,\qquad
g_{\bar \tau \tau} = g_{\tau \bar \tau} =  ({\rm Im}\, \tau)^{-2}\,.
\ee
The curvature tensor is
\be \label{curv_tensor}
R_{ijkl}=\frac{1}{2}(g_{il}g_{jk} - g_{ik}g_{jl}) \, , \quad R_{\bar \tau \tau \bar \tau \tau}=\frac{1}{2}g_{\tau \bar \tau}g_{\bar \tau \tau}
\,.
\ee
The coordinate $\tau$  thus describes the Poincare half-plane model of the 2d constant negative curvature space. 

Using the connected four-point function 
\be 
\langle \phi_{\bar \tau}(1)\phi_{\tau}(\eta)\phi_{\bar \tau}(\infty)\phi_{\tau}(0)\rangle_{c}
=({\rm Im}\,\tau)^{-4}\Bigl[ \frac{1}{(1-\eta)^{2}} + \frac{1}{(1-\bar \eta)^{2}} \Bigr]
\ee
we obtain from our general formula (\ref{eq:curvature}) 
\be
R_{\bar \tau\tau\bar \tau \tau}= -({\rm Im}\,\tau)^{-4} \int \frac{d^{2}\eta}{2\pi}
\ln|\eta| \Bigl[ \frac{1}{(1-\eta)^{2}} + \frac{1}{(1-\bar \eta)^{2}} \Bigr]
\,.
\ee
Regularizing, as prescribed, by cutting a small circle around $\eta=1$ and a large circle around the origin 
we obtain 
\be
R_{\bar \tau \tau \bar \tau \tau}=\frac{1}{2}({\rm Im}\, \tau)^{-4}
\ee 
matching (\ref{curv_tensor}).

\section{The space of conformal boundary conditions}
\setcounter{equation}{0}

We now turn to the case of boundary conformal field theories.
A boundary conformal field theory (BCFT) is
a conformal field theory 
on the disk 
with a conformally invariant boundary condition on the boundary 
circle.
As in the bulk, the BCFT's are supposed to be
unitary, with discrete spectrum, and to be invariant under the
global conformal group.

The disk can be mapped conformally to the upper 
half-plane with
the boundary becoming the projective line --- 
the real axis plus the point at infinity.
We find it convenient to calculate in the
coordinate 
$x=\tan(\theta/2)$, $-\pi \le \theta <  
\pi$,
on the projective line.
For purposes of regularization, we use the metric 
transported from the unit circle
\eq
(ds)^{2} = (d\theta)^{2} = \rho(x)^{2} (dx)^{2}\,,\qquad \rho(x) = 
\frac1{1+x^{2}}\,,
\label{eq:boundarycirclemetric}
\en
because it treats all points on the boundary uniformly, including  
the point at $x=\infty$.

We are studying smooth families of boundary CFT's for a given, fixed bulk CFT.
Such a family --- that is, a smooth family of conformal boundary 
conditions for the given CFT ---
is parameterized 
by dimensionless coupling constants $\lambda^{a}$ which couple to 
local, dimension 1 boundary 
fields $\psi_{a}(x)$ so that 
\be \label{boundary_action}
\frac{\partial}{\partial \lambda^{a}} \langle {\cal O} \rangle = 
\int\!\! dx\, \langle \psi_{a}(x) \,
{\cal O} \rangle
\ee 
where ${\cal O}$ stands for an arbitrary product of local operators.
The natural metric\footnote{
A metric on the space of not-necessarily-conformal boundary conditions
was defined in 
\cite{Moore_etal, FKg} in connection 
with the proof of the $g$-theorem \cite{AffLud,FKg},
\be 
g_{ab} = \frac{2}{\pi}\int\limits_{0}^{2\pi}\!\!d\theta\, 
\sin^{2}\left (\frac{\theta-\theta'}2\right ) \langle \psi_{a}(\theta)\psi_{b}(\theta')\rangle 
\,.
\nonumber
\ee
For conformal boundary conditions, 
this agrees with (\ref{eq:boundarymetric}).
}
on the family of BCFT's is read off from the 
two-point function
\be
\langle \psi_{a}(x_{1}) \psi_{b}(x_{2})\rangle  = g_{ab} (x_{1}-x_{2})^{-2}
\ee
or
\be
g_{ab} = \langle \psi_{a}(0) \psi_{b}(\infty)\rangle  \,,\qquad
\psi_{b}(\infty) = \lim_{x\rightarrow \infty} x^{2} \psi_{b}(x)
\,.
\label{eq:boundarymetric}
\ee

We choose a reference BCFT
satisfying 
assumptions  similar to those
made for the reference bulk CFT:\\[1ex]
\noindent {\bf Assumption 1b\;}
{\it
The $\psi_{a}(x)$ are dimension 1 boundary  fields.
}
\vskip1ex

The remaining two assumptions have to do with the 
$\psi_{a}\,\psi_{b}$ OPE,
whose singular part has the following general form
for $x>0$
\be \label{psiOPE}
\psi_{a}(x) \, \psi_{b}(0) \sim
x^{-2} g_{ab} \mathbf{1}
+ x^{-1} \sum_{\tilde c} C^{\tilde c}_{ab} \psi_{\tilde c}(0)
+ \sum_{c'} x^{\Delta_{c'}-2} C^{c'}_{ab} \psi_{c'}(0)\,.
\ee
Here, the $\psi_{\tilde c}$ are all the dimension 1 fields, which 
include our perturbations $\psi_{c}$.  The $\psi_{c'}$ are all the 
relevant boundary fields (except for the identity $\mathbf{1}$) ---
the fields of scaling dimensions $\Delta_{c'}< 1$.
The OPE for $x<0$ is, by translation invariance,
\eq
\psi_{a}(x) \, \psi_{b}(0) \sim \psi_{b}(-x) \, \psi_{a}(0) 
\,.
\en

\noindent {\bf Assumption 2b\;}
{\it
The OPE coefficients $C^{\tilde c}_{ab}$ are antisymmetric:
$C^{\tilde c}_{ab}= - C^{\tilde c}_{ba}$.
}
\vskip1ex

\noindent{\bf Assumption 3b\;} 
{\it
The OPE coefficients $C^{c'}_{ab}$ are symmetric,
$C^{\tilde c}_{ab}= C^{\tilde c}_{ba}$,
for all dimension 0 fields $\psi_{c'}$.
}
\vskip1ex

Assumptions 1b and 2b imply that the beta functions for the couplings 
$\lambda^{a}$ vanish at least through the second order.
Assumption 1b excludes boundary perturbations by derivative fields.
Assumption 3b parallels bulk assumption 3.  Dimension 0 boundary 
fields other than the identity
arise when the BCFT has degenerate ground states (described in 
string theory by Chan-Paton indices).
The dimension 0 fields act as charges which generate global 
symmetries, mixing the degenerate sectors of the BCFT.
Assumption 3b means that the perturbations commute with these charges,
that there are no boundary condition changing perturbations.
In Appendix~\ref{appendix-Cmij} we show that, if there is a charged 
perturbation, then a certain linear combination of the
$\psi_{a}$ becomes a derivative fields at first order in the symmetry breaking 
perturbation.
Therefore, as in the bulk, our assumptions systematically exclude 
derivative operators.

\section{The boundary curvature formula}

We use a hard sphere cutoff  on the 
boundary, 
renormalizing the correlation functions by minimal subtraction.
The cutoff is $d(x_{1},x_{2})>\epsilon'$
where the distance function is carried over from the unit 
circle
\be \label{proj_distance}
d(x_{1},x_{2})=|\theta_{1}-\theta_{2}|=2 \left|\tan^{-1}x_{1}-
\tan^{-1}x_{2}\right|\, . 
\ee 
Thus the cutoff can be written equivalently
\be
\left| \frac{x_1-x_{2}}{1+x_{1}x_{2}}\right|\ge \epsilon \,,
\ee
where $\epsilon = \tan \left ( {\epsilon'}/2 \right )$.
The cutoff function is
\be 
h_{\epsilon}(x_{1},x_{2})=\theta\left( 
\left | \frac{x_1 - x_2}{1+x_1 x_2} \right |- \epsilon \right) \, . 
\ee
In particular 
\be 
h_{\epsilon}(x,0)=\theta(|x|-\epsilon)\, , \qquad 
h_{\epsilon}(x, \infty)=\theta(|x|^{-1}-\epsilon)\,.  
\ee
The regulated correlation functions are
\be
\langle \psi_{a_{1}}(x_{1})\cdots \psi_{a_{N}}(x_{N}) \rangle\,
\prod_{\alpha <\beta} h_{\epsilon}(x_{\alpha},x_{\beta})
\,.
\ee

The first derivatives of the metric at $\lambda^{a}=0$ are given by
\be 
{\partial_{c}g_{ab}} = \int\!\!dx \, \langle 
\psi_{c}(x )\psi_{a}(0)\psi_{b}(\infty)\rangle \,.
\label{eq:bfirstder}
\ee 
It follows from assumption 2b
that the cutoff integral vanishes,
\eq
\int\limits_{\epsilon\le |x |\le \epsilon^{-1}}\!\!dx \, \langle 
\psi_{c}(x )\psi_{a}(0)\psi_{b}(\infty)\rangle
= 0
\,.
\en
Therefore, since we are using minimal subtraction, 
no contact terms contribute to (\ref{eq:bfirstder}).
We conclude that ${\partial_{c}g_{ab}} =0$.
The curvature tensor is given by
\be\label{Rbd}
R_{abcd} = \frac{1}{2}( \partial_{b}\partial_{c}g_{ad} - \partial_{a}\partial_{c}g_{bd} 
- \partial_{b}\partial_{d}g_{ac} + \partial_{a}\partial_{d}g_{bc})\, . 
\ee
The regularized second derivatives of the metric are
\be \label{f1}
(\partial_{b}\partial_{c}g_{ad})^{\epsilon}= \iint\!\! dx_{1}dx_{2}\, 
\langle 
\psi_{b}(x_{1})\psi_{c}(x_{2})\psi_{a}(0)\psi_{d}(\infty)\rangle_{c}H_{\epsilon}(x_{1}, x_{2}) 
\ee
where
\be 
H_{\epsilon}(x_{1}, x_{2})= h_{\epsilon}(x_{1},x_{2})h_{\epsilon}(x_{1},0)h_{\epsilon}(x_{1},\infty)
h_{\epsilon}(x_{2},0)h_{\epsilon}(x_{2},\infty)\, . 
\ee
The regularized curvature tensor $R_{abcd}^{\epsilon}$ is obtained by using the regularized  derivatives of metric (\ref{f1}) 
in (\ref{Rbd}). The curvature tensor is then obtained as 
\be 
R_{abcd}= \lim_{\epsilon \to 0} (R_{abcd}^{\epsilon} + \Delta R_{abcd}^{\epsilon}) 
\ee
where $\Delta R_{abcd}^{\epsilon}$ is the counterterm. 

We write
\eq
R^{\epsilon}_{abcd} = \frac12 \left ( \tilde R^{\epsilon}_{abcd}
- \tilde R^{\epsilon}_{abdc}
\right )
\en
with
\eq
\tilde R^{\epsilon}_{abcd}
=
(\partial_{b}\partial_{c}g_{ad})^{\epsilon} - (\partial_{a}\partial_{c}g_{bd})^{\epsilon} 
\en
which is
\eq
\label{R4pt}
\tilde R^{\epsilon}_{abcd}=\int_{\cal R^{\epsilon}}dx_{1}dx_{2}[\langle \psi_{b}(x_{1})\psi_{c}(x_{2})\psi_{a}(0)\psi_{d}(\infty)\rangle_{c} 
- \langle \psi_{b}(0)\psi_{c}(x_{2})\psi_{a}(x_{1})\psi_{d}(\infty)\rangle_{c}]
\en
where the integration region is 
\be
{\cal R^{\epsilon}}=\{ (x_{1},x_{2}):H_{\epsilon}(x_{1}, x_{2})= 1\}\, .
\label{eq:Repsilonregion}
\ee
Changing the variables of integration to $\chi=x_{2}/x_{1}$, $x=x_{1}$ and using the global conformal 
invariance of the correlation function we rewrite $\tilde R^{\epsilon}_{abcd}$ 
as
\be
\tilde R^{\epsilon}_{abcd}=\int d\chi\, F_{\epsilon}(\chi) [\langle \psi_{b}(0)\psi_{c}(\chi)\psi_{a}(1)\psi_{d}(\infty)\rangle_{c} 
+ \langle \psi_{b}(1)\psi_{c}(1-\chi)\psi_{a}(0)\psi_{d}(\infty)\rangle_{c} ]
\ee
where 
\be
F_{\epsilon}(\chi) = \Bigg[ \int_{{\cal R}^{\epsilon}_{+}(1-\chi)} - \int_{{\cal 
R}_{+}^{\epsilon}(\chi)}\Bigg] \frac{dx}{x}
\label{eq:Fepsilon}
\ee
and 
\be
{\cal R}^{\epsilon}_{+}(\chi) = \{x: x>0, (x,\chi x) \in {\cal R}^{\epsilon}\} \, .
\label{eq:Repsilonplusregion}
\ee
We show in appendix \ref{mproof} that $\tilde R^{\epsilon}_{abcd} = 
- \tilde R^{\epsilon}_{abdc}$, so
\eq
R^{\epsilon}_{abcd} = \tilde R^{\epsilon}_{abcd}\,.
\label{metric_id}
\en

We have now succeeded in expressing the regularized curvature tensor as a 
single integral.
It is straightforward but very tedious to calculate 
$F_{\epsilon}(\chi)$. The result is a piecewise continuous function
given in table \ref{table}.
Away from the singular points $\chi=0,1,\infty$
\eq
\lim_{\epsilon\rightarrow 0} F_{\epsilon}(\chi) = -\ln|1-\chi^{-1}|
\,.
\en

Next we analyze the regularization.
We write $R^{\epsilon}_{abcd}$ as the principal value regulated integral
plus an error term,
\be
R^{\epsilon}_{abcd}= R^{\rm PV}_{abcd} + E_{abcd}\, 
\ee
 where 
 \be 
 R^{\rm PV}_{abcd}=\int d\chi F^{\rm PV}_{\epsilon}(\chi) [\langle \psi_{b}(0)\psi_{c}(\chi)\psi_{a}(1)\psi_{d}(\infty)\rangle_{c} 
+ \langle \psi_{b}(1)\psi_{c}(1-\chi)\psi_{a}(0)\psi_{d}(\infty)\rangle_{c} ]
 \ee
 with 
 \be 
 F^{\rm PV}_{\epsilon} =  -\ln|1-\chi^{-1}| \, 
 \theta(\chi - \epsilon^{2})\theta(1-\chi - \epsilon^{2})\theta(\chi^{-1}-\epsilon^{2}) 
\label{eq:FPV}
 \ee
 and 
 \be \label{Edef}
 E_{abcd}=\int d\chi \Delta F_{\epsilon}(\chi)[\langle \psi_{b}(0)\psi_{c}(\chi)\psi_{a}(1)\psi_{d}(\infty)\rangle_{c} 
+ \langle \psi_{b}(1)\psi_{c}(1-\chi)\psi_{a}(0)\psi_{d}(\infty)\rangle_{c} ]
 \ee
where
\eq
\Delta F_{\epsilon}(\chi) = F_{\epsilon}(\chi) - F^{\rm PV}_{\epsilon}(\chi)
\label{eq:DeltaFepsilon}
\,.
\en
The function $\Delta F_{\epsilon}(\chi)$ is also given in table \ref{table}.
 
The next step is to show that the combination of the error term 
$E_{abcd}$ and the renormalization counterterm  $\Delta R_{abcd}^{\epsilon}$
gives the minimal subtraction for principal value regularization.
That is,
 \be
 E_{abcd} + \Delta R_{abcd}^{\epsilon} = - (R^{\rm PV}_{abcd})_{\rm sing} + r(\epsilon)
 \ee 
 where $(R^{\rm PV}_{abcd})_{\rm sing}$ is the singular part of $R^{\rm PV}_{abcd}$ 
and 
 $r(\epsilon)\to 0$ as $\epsilon \to 0$.
 
The singularities of the four-point function are found using the OPE (\ref{psiOPE}). 
For $\chi\rightarrow 0$,
\aeq{
\expvalc{\psi_{b}(0) \psi_{c}(\chi) \psi_{a}(1) \psi_{d}(\infty)}
&\sim
\chi^{-1} \sum_{\tilde e} C^{\tilde e}_{cb} C_{da\tilde e}
\nonumber\\ & \quad{}
+
\sum_{e'} |\chi|^{\Delta_{e'}-2}
\left [ C^{e'}_{cb} \theta(\chi)
+C^{e'}_{bc}  \theta(-\chi) \right ] C_{dae'}\,,
\\[1ex]
\expvalc{\psi_{b}(0) \psi_{c}(1-\chi) \psi_{a}(1) \psi_{d}(\infty)}
&\sim
\chi^{-1} \sum_{\tilde e} C^{\tilde e}_{ca} C_{db\tilde e}
\nonumber\\ & \quad{}
+
\sum_{e'} |\chi|^{\Delta_{e'}-2}
\left [ C^{e'}_{ac}  \theta(\chi) +C^{e'}_{ca} \theta(-\chi)
\right ] C_{bde'}\,.
}
For $\chi\rightarrow \infty$
\aeq{
\expvalc{\psi_{b}(0) \psi_{c}(\chi) \psi_{a}(1) \psi_{d}(\infty)}
&\sim
\chi^{-1} \sum_{\tilde e} C_{dc\tilde e} C^{\tilde e}_{ab}
\nonumber\\ & \quad{}
+
\sum_{e'} |\chi|^{-\Delta_{e'}}
\left [ C_{dce'} \theta(\chi)
+C_{cde'}  \theta(-\chi) \right ] C^{e'}_{ab} \, .
}
We have defined the OPE coefficients with lowered indices by
\be 
C_{ab\tilde c}=\langle\psi_{a}(1)\psi_{b}(0)\psi_{\tilde c}(\infty)\rangle \,, \quad 
C_{abc'}=\langle\psi_{a}(1)\psi_{b}(0)\psi_{c'}(\infty)\rangle \,.
\ee
Using these expressions for the singular parts of the four-point 
function, we obtain
\be 
 (R^{\rm PV}_{abcd})_{\rm sing}=\sum_{c'}
\left [ -\ln (\epsilon^{2})\frac{(\epsilon^{2})^{\Delta_{c'}-1}}{1-\Delta_{c'}} 
- \frac{(\epsilon^{2})^{\Delta_{c'}-1}}{(1-\Delta_{c'})^{2}}
\right ]
K^{c'}_{abcd}
\ee
 where 
\be\label{K}
K^{c'}_{abcd}=  C^{c'}_{(ac)}C_{(bd)c'}-C^{c'}_{(ad)}C_{(bc)c'}\, , \qquad                                   
C^{c'}_{(ac)}=C^{c'}_{ac} + C^{c'}_{ca}\, .
\ee
Note that the dimension 1 fields make no contribution, because of the 
principal value regularization.
We notice in calculating $(R^{\rm PV}_{abcd})_{\rm sing}$ that there 
is no contribution from $\chi=\infty$ because of the factor
$-\ln|1-\chi^{-1}|$ in (\ref{eq:FPV}).

 
We next discuss the renormalization counterterm. In our regularization scheme the divergences from 
a pair of colliding insertions are obtained from the OPE (\ref{psiOPE}) 
\be 
 \int dx_{2}\, \psi_{a}(x_{2})\psi_{b}(x_{1})h_{\epsilon}(x_{1},x_{2}) 
 \sim \sum_{c'}\frac{\epsilon^{\Delta_{c'}-1}}{1-\Delta_{c'}}C^{c'}_{(a,b)}
\rho(x_{1})^{1-\Delta_{c'}}\psi_{c'}(x_{1})
\ee
where $\rho(x)$ is the scale factor of the metric on the boundary 
given in (\ref{eq:boundarycirclemetric}).
This implies a counterterm for the action
\eq\label{dS}
\Delta S = \int dx \, \rho(x)\,
\frac12 \sum_{c'} 
\frac{\epsilon^{\Delta_{c'}-1}}{1-\Delta_{c'}}
C^{c'}_{(ab)} \lambda^{a}\lambda^{b} \, \rho(x)^{-\Delta_{c'}} \psi_{c'}(x)
\,.
\en
It is a standard calculation to find the contribution to the 
four-point functions of this counterterm to the action.
We find 
\eq\label{dRabcd}
\Delta R_{abcd}
=
\sum_{c'} \frac{\epsilon^{\Delta_{c'}-1}}{1-\Delta_{c'}}
\int_{0}^{\infty} dx\; (1+x^{2})^{\Delta_{c'}-1} x^{-\Delta_{c'}}
K^{c'}_{abcd}
+  \sum_{c'} \left ( \frac{\epsilon^{\Delta_{c'}-1}}{1-\Delta_{c'}} 
\right )^{2}
\left (
-K^{c'}_{abcd}
\right )
\en
where $K^{c'}_{abcd}$ is given in (\ref{K}).

 The error term $E_{abcd}$ defined in (\ref{Edef}) can be evaluated explicitly up to 
 terms tending to zero as $\epsilon \to 0$. We find
\begin{multline}
E_{abcd}=
\sum_{c'}\ln (\epsilon^{2})\frac{(\epsilon^{2})^{\Delta_{c'}-1}}{1-\Delta_{c'}} 
K^{c'}_{abcd}
+ \sum_{c'}\frac{(\epsilon^{2})^{\Delta_{c'}-1}}{(1-\Delta_{c'})^{2}}
2K^{c'}_{abcd} 
\\
+  \sum_{c'}\frac{\epsilon^{\Delta_{c'}-1}}{1-\Delta_{c'}}
\int_{2}^{\infty}
du \; \left [
-\partial_{u} A(u^{-2}) + 2 u^{-1}
\right ]u^{\Delta_{c'}-1}
\left (
-K^{c'}_{abcd}
\right )
\label{eq:Eerrorterm}
\end{multline}
 where 
 \be 
 A(u^{-2})=-2\ln\left(\frac{1}{2}+\sqrt{\frac{1}{4}-\frac{1}{u^{2}}}   \right)\, . 
 \ee
 The details of this computation are put into appendix \ref{Edetails}.
The key point is that there are only singular terms  in the limit 
$\epsilon\rightarrow 0$, no finite terms.
Since the $\Delta R_{abcd}$ counterterm also contains only singular terms,
we can conclude that
 \be 
 E_{abcd} + \Delta R_{abcd} = - (R^{\rm PV}_{abcd})_{\rm sing}
 \ee
up to terms vanishing in the limit $\epsilon \to 0$.
We can verify this equation explicitly using the identity 
 \be 
 \int_{2}^{\infty}
du \; \left [
-\partial_{u} A(u^{-2}) + 2 u^{-1}
\right ]u^{\Delta -1} = - \int\limits_{0}^{\infty}dx \, 
(1+x^{2})^{\Delta-1}x^{-\Delta} \,.
 \ee
 We thus arrive at the following formula for the curvature 
\aeq{
\label{Rfinal}
 R_{abcd} &= {\rm RV}\! \int\!\!\! d\chi\, (-\ln|1-\chi^{-1}|)
 [\langle 
 \psi_{b}(0)\psi_{c}(\chi)\psi_{a}(1)\psi_{d}(\infty)\rangle_{c} 
 \nonumber\\
 &\qquad\qquad\qquad\qquad\qquad\qquad{}
+ \langle \psi_{b}(1)\psi_{c}(1-\chi)\psi_{a}(0)\psi_{d}(\infty)\rangle_{c} ]
}
where the integral near $\chi=0,1$ is taken with principle value 
regularization and minimal subtraction. As we remarked before, no regularization is needed at $\chi=\infty$. 

It should be noted that even when there are no relevant 
operators in the OPE,
the integral in the curvature formula is 
still conditionally convergent around $\chi=0,1$ in general.
The principal value prescription is still needed.   

Changing integration variable to  $\eta=1-\chi^{-1}$ and making a 
conformal transformation of the four-point functions,
we obtain the boundary curvature formula stated in the Introduction
\aeq{
R_{abcd} &= {\rm RV} \! \int\limits_{-\infty}^{\infty} d\eta\; (-\ln 
|\eta|)\;
\left [
\expvalc{\psi_{a}(1)\psi_{b}(\eta)\psi_{c}(\infty)\psi_{d}(0)}
\right .
\nonumber\\
&\qquad\qquad\qquad\qquad\qquad
\left . {}+\expvalc{\psi_{a}(0)\psi_{b}(1-\eta)\psi_{c}(\infty)\psi_{d}(1)}
\right ]
\,.
\label{Rfinal2}
}
In this formula, no regularization is needed at $\eta=1$.
The change of variable  $\eta=1-\chi^{-1}$ does not manifestly 
preserve the principal value regularization, so care is needed to 
check that the regularization is in fact preserved, given our assumptions.

As in the bulk, the boundary curvature formula depends only on 
the correlation functions at finite separation, so is coordinate 
independent.


\section{D0 branes on group manifolds}\label{sec_D0}
\setcounter{equation}{0}
As a check of the boundary curvature formula (\ref{Rfinal2}) we will consider the example of D0 brane boundary conditions 
on group manifolds. 
The bulk CFT is a WZW theory at level $k$ for
a semisimple compact Lie group $G$.
We pick a basis in the Lie algebra so that the corresponding currents 
$J^{a}(z)$ satisfy the OPE
\be 
J^{a}(z)J^{b}(w) \sim \frac{k \delta^{ab}}{(z-w)^{2}} + \frac{i{f^{ab}}_{c}J^{c}(w)}{z-w} + \dots
\ee 
where $f^{ab}_{c}$ is the totally antisymmetric tensor of  the Lie group structure constants.
As a reference boundary condition we take the D0 brane located at the identity element.
The boundary condition on a half plane glues the left and right 
components of the currents at the boundary as
 $J^{a}(x ) = \bar J^{a}(x )$. As shown in \cite{RS} the boundary perturbation 
 \be
 \langle \exp(\int\!\! dx \, \sum_{a} \lambda^{a}J^{a}(x )) \dots \rangle 
 \ee
 is exactly marginal for all values of the couplings $\lambda^{a}$.
 The $\lambda^{a}$ parameterize the position $g(\lambda)$ of the D0 brane in the group manifold $G$. 
 The corresponding boundary   condition is
 \be
 J^{a}(x ) = ({\rm Ad }_{g} \bar J)^{a}(x ) 
 \ee 
 where ${\rm Ad}_{g}$ stands for the adjoint action of $G$ on its Lie 
 algebra.
 Since the moduli space is a homogeneous space it suffices to compute the curvature at a single point.
We will first compute the curvature in terms of double integrals of
distributional four-point functions, as in \cite{Kutasov:1988xb}.
Then we check that the result agrees with our formula (\ref{Rfinal2}).

We will be calculating first and second derivatives of the metric 
which is given by the two-point function at finite separation
\eq
\expval{J^{c}(x_{3}) J^{d}(x_{4})} = \frac{g_{cd}}{ (x_{3}-x_{4})^{2}}
\,.
\en
The distributional correlation functions on the boundary are defined in 
Appendix~\ref{app1}.
To find the first derivatives of the metric,
we integrate the three-point function (\ref{3pt2}),
\be 
\int\limits_{-\infty}^{+\infty}\!\!dx _{1}\, \langle J^{a}(x 
_{1})J^{c}(x _{3})J^{d}(x _{4})\rangle 
= 0
\ee
at finite separation.
Thus, in our coordinates,
\be 
{\partial_{a}g_{cd}}_{/\lambda^{a}=0}=0
\ee
 so we can use formula (\ref{Rbd}) to compute the curvature at the origin\footnote{It is easy 
 to see that any regularization of the three-point function of currents
 which preserves the group symmetry will have the same property.}. 
 
Now we have to calculate the second derivatives of the metric.
Integrating  the distributional four-point 
 function (\ref{4pt_distr}) once, we get 
\aeq{
& \int\limits_{-\infty}^{+\infty}\!\! dx _{1}\, \langle J^{a}(x 
 _1)J^{b}(x _2)J^{c}(x _3)J^{d}(x _4)\rangle \nonumber \\
& \qquad\qquad\qquad     = \frac{k\pi^{2}}{3}{f^{ac}}_{e}f^{bed}\Bigl(  2\delta(x _{23})\Bigl[\frac{1}{x _{24}^{2}}  \Bigr]
   -\delta(x _{34})\Bigl[\frac{1}{x _{24}^{2}}  \Bigr] - \delta(x _{24})\Bigl[\frac{1}{x _{23}^{2}}  \Bigr]\Bigr) \nonumber\\
&  \qquad\qquad\qquad \quad     + \frac{k\pi^{2}}{3}{f^{ad}}_{e}f^{bce}\Bigl( -2\delta(x _{24}) \Bigl[\frac{1}{x _{23}^{2}}  \Bigr] 
   + \delta(x _{34})\Bigl[\frac{1}{x _{23}^{2}}  \Bigr] + \delta(x 
   _{23})\Bigl[\frac{1}{x _{43}^{2}}  \Bigr]   \Bigr)
}
where the square brackets stand for the distributional regularization
\eq
\left [ \frac1{x^{2}} \right ]  = -\partial_{x} \mathrm{PV} \left ( \frac1{x}\right )
\,.
\en
Integrating one more time we obtain 
\eq
 \int\limits_{-\infty}^{+\infty}\!\! dx _{1}\int\limits_{-\infty}^{+\infty}\!\! dx _{2}\,
 \langle J^{a}(x _1)J^{b}(x _2)J^{c}(x _3)J^{d}(x _4)\rangle 
 = \frac{k\pi^{2}}{3}\Bigl[\frac{1}{x _{34}^{2}}\Bigr]( 
   {f^{ac}}_{e}f^{bed} + {f^{ad}}_{e}f^{bec}   )
\en   
so
\eq
\partial_{a}\partial_{b}g_{cd}  = \frac{k\pi^{2}}{3}( 
   {f^{ac}}_{e}f^{bed} + {f^{ad}}_{e}f^{bec}   )\,.
   \label{eq:secondderivativesbdry}
\en
   From this expression it is easy to see that 
   \be 
   {\partial_{a}\partial_{b}g_{cd}}_{/\lambda^{a}=0}={\partial_{c}\partial_{d}g_{ab}}_{/\lambda^{a}=0} 
   \ee
   and 
   \be 
   {[\partial_{a}\partial_{b}g_{cd} + \partial_{a}\partial_{c}g_{db} + 
   \partial_{a}\partial_{d}g_{bc}]}_{/\lambda^{a}=0} = 0
   \ee
  implying
   that the distributional correlators defined in Appendix \ref{app1} correspond to Riemann 
   normal coordinates at the origin.
From (\ref{eq:secondderivativesbdry}) we obtain 
   \be \label{result_direct}
   R_{abcd}=k\pi^{2}{f^{ab}}_{e}f^{ced}\,.
   \ee 
The Killing metric $g^{\rm Killing}_{ab}$ on the group manifold has 
curvature tensor
   \be
   R^{\rm Killing}_{abcd}=\frac{1}{4}{f^{ab}}_{e}f^{ced}
   \ee
so the metric on the space of conformal boundary conditions
is
\be
   g_{ab} = 4\pi^{2}k \,g^{\rm Killing}_{ab}\, .
   \ee 
Next we check that 
our general curvature formula (\ref{Rfinal2})
gives the same result.
The four-point function is
   \be
\langle 
J^{b}(\infty)J^{c}(\eta)J^{a}(0)J^{d}(1)\rangle_{c}=-\frac{k}{1-\eta}{f^{ba}}_{e}f^{ced} 
+\frac{k}{\eta}{f^{bd}}_{e}f^{cae}
\,.
\ee
Substituting into (\ref{Rfinal2}) gives
\be \label{R_int}
R_{abcd}= -2k\Bigl[{f^{bd}}_{e}f^{cae} I_{1} + {f^{ba}}_{e}f^{ced} I_{2}  \Bigr]
\ee
where 
\be 
I_{1}=\mathrm{PV} \int\limits_{-\infty}^{\infty}\!\!d\eta 
\;\frac{\ln|\eta|}{\eta} = 0\,
\ee
\be
I_{2}=-\mathrm{PV}\int\limits_{-\infty}^{\infty}\!\!d\eta\;  
\frac{\ln|\eta|}{1-\eta} = \frac{\pi^{2}}2\,,
\ee
so  (\ref{R_int}) agrees
with the direct computation (\ref{result_direct}).

\section{Curvature formula and string theory effective action} 
Here we show that the curvature tensor (\ref{eq:curvature}) appears in the low 
energy action for massless scalars in string theory.

Suppose we have a CFT with integer central 
charge $c\le 24$.  The tensor product with $d=26-c$ free bosons $X^{\mu}$ is a bosonic 
string background. The massless scalar vertex operators are 
\be
V_{i}=:\! e^{iP\cdot X}\!:\, \phi_{i}\, ,  \quad P^{2}=0\, .
\ee
The Virasoro-Shapiro four-point amplitude for the massless scalars is 
\eq
\delta^{d}\left ( \sum P_{\alpha}\right ) 
\mathcal{A}_{i_{1}i_{2}i_{3}i_{4}}^{(4)}(s,t,u)
=
\int \frac{d^{2}\eta}{2\pi} \;
\expvalc{V_{i_{1}}(1)\;V_{i_{2}}(\eta)\;V_{i_{3}}(\infty)\;V_{i_{4}}(0)}
\en
with on-shell condition $s+t+u=0$.
Substituting for the $V_{i}$ and evaluating the free boson 
correlation functions, we obtain
\aeq{
\mathcal{A}_{i_{1}i_{2}i_{3}i_{4}}^{(4)}(s,t,u) &=
\Bigl(
\frac{tu\left(t+2\right)\left(u+2\right)}{8s\left(s+2\right)}g_{i_{1}i_{2}}g_{i_{3}i_{4}}
+\frac{su\left(s+2\right)\left(u+2\right)}{8t\left(t+2\right)}g_{i_{1}i_{3}}g_{i_{2}i_{4}}\nonumber\\
 &\qquad{} +\frac{st\left(s+2\right)\left(t+2\right)}{8u\left(u+2\right)}g_{i_{1}i_{4}}g_{i_{2}i_{3}}
\Bigr)
F(s,t,u)\nonumber
\\
&\quad
{}+
\int\!\! \frac{d^{2}\eta}{2\pi} \;
|\eta|^{- t} |1-\eta|^{- s}\;
\expvalc{\phi_{i_{1}}(1)\;\phi_{i_{2}}(\eta)\;\phi_{i_{3}}(\infty)\;\phi_{i_{4}}(0)}
\label{eq:VS4pt}
}
where 
\eq
F(s,t,u) =\frac{
\Gamma\left(1-\frac12 t\right)
\Gamma\left(1-\frac12 s\right) 
\Gamma\left(1-\frac12 u\right) 
}
{
\Gamma\left(2+\frac12 t\right)
\Gamma\left(2+\frac12 s\right) 
\Gamma\left(2+\frac12 u\right) 
}
\,.
\en

The usual assumption is that the low energy string 
scattering amplitudes come from an effective $d$-dimensional
field theory action.
The part that describes the self-interactions of the
massless scalar fields $\Phi^{i}(X)$ is
the $d$-dimensional non-linear sigma model 
\be 
S_{\mathit{eff}} = \int d^{d}X \; 
\frac12 g_{ij}(\Phi)\partial_{\mu}\Phi^{i}\partial^{\mu}\Phi^{j}
\ee
where the $\Phi^{i}$ are coordinates on the space of CFT's and
$g_{ij}$ is the Zamolodchikov metric.
As far as we know, this has never been proved.
Accepting the assumption, 
the low energy limit of the four-point scattering amplitude due to
self-interactions can be
calculated by expanding
the metric in Riemann normal coordinates,
\be
S_{\mathit{eff}} =
\int d^{d}X \; 
\left (
\frac12 \delta_{ij}\partial_{\mu}\Phi^{i}\partial^{\mu}\Phi^{j}
-\frac{1}{3}
R_{ikjl}\Phi^{k}\Phi^{l}\partial_{\mu}\Phi^{i}\partial^{\mu}\Phi^{j}
+ \cdots
\right )
\ee
giving
\aeq{ \label{VS_exp}
\mathcal{\tilde A}_{i_{1}i_{2}i_{3}i_{4}}^{(4)}(s,t,u) &=
t R_{i_{1}i_{4}i_{3}i_{2}} +u  R_{i_{1}i_{3}i_{4}i_{2}} 
+O(s^{2},t^{2},st)
\,.
}
We can compare with the string theory amplitude (\ref{eq:VS4pt})
if we drop the first three terms, which in the low energy limit come 
from tachyon and graviton exchange.  We then formally obtain our 
curvature formula~(\ref{eq:curvature}).
We say `formally', because we have not addressed the issues of 
regularization.  Assuming that those issues can be handled, our 
proof of the curvature formula becomes a point of support for the 
effective action assumption.

\section{Discussion}
\setcounter{equation}{0}
We conclude with brief remarks on two topics:
the possibilty of a general bound on the sectional curvature
and the extension of the curvature formula to neighborhoods 
of CFT's with continuous symmetries.

Formulas (\ref{eq:curvature}), (\ref{eq:bcurvature}) express the 
curvature of the space of CFTs in terms 
of intrinsic CFT quantities --- the four-point correlation functions. 
The correlation functions of a CFT satisfy reflection positivity, 
conformal invariance, and crossing symmetry.
One might hope to use these properties to say something 
about the geometry of the space of CFTs.  

One possibility is that reflection positivity of the four-point
functions implies a bound on the sectional curvatures.  
Let directions $i=1,2$ be mutually orthogonal.
From the curvature formula (\ref{eq:curvature}) we can write
the sectional curvature in the 1-2 plane as
\be 
R_{2112}=\lim_{\epsilon \to 0} \int\limits_{|z-1|>\epsilon,\, \epsilon<|z|<\epsilon^{-1}}\frac{d^{2}z}{2\pi} 
\ln|1-z| \langle \phi_{2}(\infty) \phi_{1}(z) \phi_{1}(1)\phi_{2}(0) \rangle_{c} 
\ee 
where
\be 
\langle \phi_{2}(\infty) \phi_{1}(z) \phi_{1}(1)\phi_{2}(0) \rangle_{c} = 
\langle \phi_{2}(\infty) \phi_{1}(z) \phi_{1}(1)\phi_{2}(0) \rangle - |1-z|^{-4}\, . 
\ee
For simplicity we have assumed no relevant operators.  We have chosen
this form of the curvature formula in order that the four-point 
function have the form appropriate for reflection positivity under the 
reflection $z\rightarrow 1/\bar z$ of the radial quantization.
The full four-point function satisfies reflection positivity, but 
the connected four-point function
does not, because of the subtraction.
We have not managed to find a way around this obstacle.
The logarithm in the integrand is another potential difficulty, but 
one might hope to get around it by using global conformal 
transformations.


Next, we discuss the cases that our curvature formula does not cover
--- the neighborhoods of CFT's with continuous symmetries.  To handle
these cases, one would have to relax our
assumptions~\ref{assumption:dim2} and
~\ref{assumption:nochiralcurrents}.  At the symmetry point, one would
have to allow for the conserved currents to occur in the OPE's of the
perturbations $\phi_{i}$.  As discussed in
appendix~\ref{appendix-Cmij}, this would imply that some linear
combinations of the $\phi_{i}$ become total derivatives away from the
symmetry point.  Therefore to cover the neighborhood of the symmetry
point, one must allow from the start for perturbations that are total
derivatives.
To derive the curvature formula at the symmetry point, one would have 
to deal with the logarithmic divergence in the integral over $\eta$ 
due to the occurrence of the current in the intermediate channels.
One would also have to deal with the effects of the current on the 
conformal transformation properties of the regularization.
To allow for total derivative perturbations $\phi_{i}$, one will face 
further technical complications stemming from their conformal dimensions being 
different from two.

\begin{center}
{\bf \large Acknowledgments}
\end{center}
We thank D. Kutasov and E. Rabinovici for correspondence and discussion.
The work of D.F. was supported by the Rutgers New High Energy Theory Center
and by U.S. Department of Energy grant DE-FG02-12ER41813. 
A.K. acknowledges the support of the STFC grants ST/G000514/1 ``String Theory Scotland''
and ST/J000310/1 ``High energy physics at the Tait Institute''.
D.F. thanks Heriot-Watt University for hospitality during an intermediate stage of the project.
A.K. thanks the Natural Science Institute of the University of Iceland for 
hospitality during the initial and final stages of this project.


\section*{Appendices}
\renewcommand{\theequation}{\thesection.\arabic{equation}}
\appendix
\section{Bulk and boundary marginally redundant operators
\label{appendix-Cmij}}
\setcounter{equation}{0}

In this appendix we elaborate on the meaning of 
assumptions~\ref{assumption:nochiralcurrents} and~3b.
We show that, for a CFT with continuous symmetry and charged 
perturbations $\phi_{i}$, some linear combinations of the $\phi_{i}$ 
become redundant, at first order in the perturbation.
We also show that the trace anomaly will contain a current term.
We give analogous results for the boundary case.




We consider a reference CFT that satisfies assumptions
\ref{assumption:dim2} and \ref{assumption:ope} but not the assumption
\ref{assumption:nochiralcurrents}.  Assumption \ref{assumption:ope} in
particular excludes the dimension 2 current-current primaries $:\!J_{m}\bar
J_{n}\!:$ from the $\phi_{i}\phi_{j}$ OPE. This implies that
only holomorphic or only
antiholomorphic currents appear in this
OPE.
The situation excluded by assumption \ref{assumption:nochiralcurrents}
is therefore a reference CFT with perturbations charged under a 
chiral symmetry group\footnote{We thank D.~Kutasov for a comment
clarifying the point that the marginal couplings must develop a 
non-zero beta function if both holomorphic and anti-holomorphic 
currents are present in their OPE's.
In~\cite{Kutasov:1988xb}, it was claimed that conformal
invariance is broken away from the symmetry point when
dimension one currents are present in the OPE of the perturbing fields.
Presumably, it was implicitly assumed
that both chiral and antichiral conserved
currents occur in the OPE.}.
Without loss of generality we restrict ourselves to the situation when the  $\phi_{i}\phi_{j}$
OPE includes 
the holomorphic currents $J_{m}(z)$ and no relevant operators,
\eq
\phi_{i}(z)\,\phi_{j}(0) \sim \frac{1}{z\bar z^{2}}C^{m}_{ij} J_{m}(z) 
\,.
\en
Assuming a real basis in the space of  currents we normalize them as 
\be
\langle J_{m}(z) J_{n}(w)\rangle = - \frac{\delta_{mn}}{(z-w)^{2}}
\ee
so that the OPE coefficients $C^{m}_{ij}$ are real and satisfy
\be
C^{m}_{ij}= -C_{mij}\, , \qquad C^{m}_{ij}=-C^{m}_{ji}\, .
\ee

The  three-point functions are
\be
\expvalc{\phi_{i}(z_{1}) \, \phi_{j}(z_{2}) \, J_{m}(z)}
= 
C_{mij}
|z_{1}-z_{2}|^{-4} (z_{1}-z_{2}) (z_{1}-z)^{-2} (z_{2}-z)^{-2}\,.
\ee

In the reference CFT,
\eq
\langle \bar T(\bar w)J_{m}(z',\bar z')\phi_{i}(z,\bar z)\rangle = 0
\,.
\en
Now we perturb by $\lambda^{j}\phi_{j}$.  At first order,
this three-point function becomes
\aeq{
\label{defTb}
\langle \bar T(\bar w)J_{m}(z',\bar z')\phi_{i}(z,\bar 
z)\rangle_{1}&=
\int\!\! \frac{d^{2}\xi}{2\pi} \langle \lambda^{j}\phi_{j}(\xi, \bar \xi) \bar T(\bar w) J_{m}(z',\bar z')
\phi_{i}(z, \bar z)\rangle  \nonumber\\ 
&=\frac{C_{mij}\lambda^{j}}{(z'-z)(\bar w - \bar z)^{2}}\int\!\! \frac{d^{2}\xi}{2\pi}
\frac{1}{(z'-\xi)(z-\xi)(\bar w - \bar \xi)^{2}}\nonumber\\
&=\frac{1}{2}C_{mij}\lambda^{j} \frac{1}{(z'-z)^{2}(\bar w-\bar z)^{2}}
\Bigl[ \frac{1}{\bar w - \bar z} - \frac{1}{\bar w - \bar z'}  \Bigr]\, .
}
We see that $\bar T$ remains anti-holomorphic, so local conformal 
invariance is unbroken.
Moreover, the correlation function decays as $\bar w^{-4}$ so global 
convormal invariance also remains unbroken.
From the $(\bar w -\bar z')^{-1}$ singularity we obtain 
\be\label{red_1}
\partial_{\bar z}J_{m}(z,\bar z) = -\frac{1}{2}C_{mij}\lambda^{j}\phi_{i}(z,\bar z)
\ee
which means that, for every symmetry broken by the perturbation, 
there is a redundant field,
given by the right hand side.
Using (\ref{defTb}) and the Ward identity for the stress-energy tensor  we find a term in the trace anomaly  
\be
\Theta(z,\bar z) \sim \lambda^{j}C_{mij}\partial_{\bar z} \lambda^{i}J_{m}(z,\bar z)
\,.
\ee
Comparing to the general expression (\ref{eq:exptheta}) for the trace 
anomaly, we see that the coefficients $C^{m}_{i}(\lambda)$
satisfy
$\partial_{i}C^{m}_{j}(0)=C_{ij}^{m}$.
We remark that the anomalous dimensions of 
the redundant operators come from this term in the trace anomaly, not 
from the beta function, which is zero.
Explicitly, the scaling dimension matrix for the $J_{m}$ is
\eq\label{anom_dims}
\Delta^{m}_{n} = \delta^{m}_{n} - \frac 14
C_{nij} \lambda^{j} {C^{mi}}_{k}\lambda^{k}
\,
\en
through the second order in the couplings. 


A simple example is given by the $c=1$ gaussian model at the 
self-dual point, which is the $SU(2)$ WZW model with $k=1$.
This  example and more general toroidal examples were discussed in \cite{Moore} (see section 9 in particular). 
We take as perturbations the fields $\phi_{i}=J_{i}(z)\bar J_{3}(\bar 
z)$, $i=1,2,3$.  The symmetry currents are $J_{i}(z)$.  The field 
$\phi_{3}$ is the perturbation which changes the radius of the free 
boson in the gaussian model.  Any perturbation 
$\lambda^{i}\phi_{i}$ can be rotated by the $SU(2)$ symmetry to 
to a perturbation by $\phi_{3}$ only, so all the perturbations 
$\lambda^{i}\phi_{i}$ preserve conformal invariance and are 
equivalent to a gaussian model away from the self-dual point.

For concreteness, consider a perturbation by $\phi_{3}$.
Let $X_{L}(z)$ and $X_{R}(\bar z)$ be the chiral parts of the free 
boson field normalized as 
\be
\langle X_{L}(z)X_{L}(w)\rangle = -\ln(z-w)\, , \qquad 
\langle X_{R}(\bar z)X_{R}(\bar w)\rangle = -\ln(\bar z-\bar w)\, . 
\ee
The current $J_{3}(z)$ is 
\be
J_{3}(z)=-\partial X_{L}(z)\, . 
\ee
The spin 1 fields $J_{1}$ and $J_{2}$ are given in terms of 
exponentials of the free boson
\bea
&& J_{1}(z,\bar z)=i\sqrt{2}:\!\cos(Q_{L}X_{L}(z)+Q_{R}X_{R}(\bar z))\!:\,\, ,  \nonumber\\
&& J_{2}(z,\bar z) = -i\sqrt{2} :\!\sin(Q_{L}X_{L}(z)+Q_{R}X_{R}(\bar z))\!:
\eea
where 
\be
Q_{L}=\frac{1}{\sqrt{2}}(R + \frac{1}{R}) \, , \quad Q_{R}=\frac{1}{\sqrt{2}}(R - \frac{1}{R})
\ee
with $R=1$ corresponding to the self dual radius. We further identify 
\be
\phi_{i}(z,\bar z) = -:\! J_{i}\bar \partial X_{R}\!:(z,\bar z) \, . 
\ee
The OPE coefficients $C_{mij}$ at the self dual point are 
\be
C_{mij}=\sqrt{2}\epsilon_{mij}\, . 
\ee
With the perturbation away from the self-dual radius, the fields $J_{1},J_{2}$ stop being holomorphic.
Their divergences become proportional to the fields $\phi_{1}$ and $\phi_{2}$
so that the latter are now redundant.
Explicitly we have
\bea 
&&\bar \partial J_{1} = -i\sqrt{2}Q_{R}:\!\sin(Q_{L}X_{L}(z)+Q_{R}X_{R}(\bar z))\!:\bar \partial X_{R} 
= -Q_{R}\phi_{2}\\
&& \bar \partial J_{2} = -i\sqrt{2}Q_{R}:\!\cos(Q_{L}X_{L}(z)+Q_{R}X_{R}(\bar z))\!:\bar \partial X_{R}
=Q_{R}\phi_{1}
\eea
which matches with formula (\ref{red_1}) upon identifying  $\lambda_{3}=R-R^{-1}$.
The  conformal 
dimensions of fields $J_{1}, J_{2}$ become  
\be
\Delta = \frac{1}{4}\left(R + \frac{1}{R}\right)^{2}\, , \bar \Delta = \frac{1}{4}\left(R - \frac{1}{R}\right)^{2}
\ee
which agrees with the general formula  (\ref{anom_dims}). 

For a general perturbation we have a family of CFT's parametrized by the 
$\lambda^{i}$.
The group $SU(2)$ acts on the acts on this family.
The point $\lambda^{i}=0$ is a fixed point of the action, so the 
group is a symmetry group of that CFT.  Away from the fixed point, 
only a $U(1)$ subgroup leaves the CFT fixed.  The full $SU(2)$ group 
generates an  $SU(2)/U(1)$ equivalence class.  The redundant fields  
are the perturbations within the equivalence class.
The situation for a general symmetry $G$ is the same. 

There are analogous phenomena in boundary CFT's.
Let $\psi_{a}(x)$ be dimension 1
boundary fields and $\chi_{m}(x)$ be dimension 0 boundary fields.
Suppose the dimension 0 fields appear in the OPE's of the dimension 1 
fields,
\be 
\psi_{a}(x)\psi_{b}(0) \sim \frac{1}{x^{2}} C^{m}_{ab} \chi_{m}
\ee 
We normalize the fields so that 
\be 
\langle \psi_{a}(x)\psi_{b}(0)\rangle = \frac{1}{x^{2}}\delta_{ab}\, , \qquad 
\langle \chi_{m}(x) \chi_{n}(0)\rangle =  \delta_{mn}
\,.
\ee
The above OPE's imply the commutation relations
\be 
[\chi_{m}, \psi_{a}(x)]=({C_{ma}}^{b}-{{C_{m}}^{b}}_{a})\psi_{b}(x)\, . 
\ee
The $\chi_{m}$ are Chan-Paton charge operators.
The matrices ${C_{ma}}^{b}-{{C_{m}}^{b}}_{a}$ give the charges of the $\psi_{a}$.
To first order in the perturbation $\lambda^{b}\psi_{b}$,
\aeq{
\label{Tpch}
\langle T(z)\psi_{a}(x')\chi_{m}(y)\rangle_{1}
&=\lambda^{b}\int\!\! dx \, \langle T(z)\psi_{b}(x )\psi_{a}(x')\chi_{m}(y)\rangle \nonumber\\
& = \lambda^{b}({C_{mb}}^{a}-{{C_{m}}^{a}}_{b})\Bigl[ \frac{1}{(z-x')^{3}} + \frac{1}{(z-x')^{2}(z-y)}\Bigr]\, . 
}
This gives the OPE in the perturbed theory
\eq
T(z)\chi_{m}(0) \sim \frac{1}{z}\lambda^{b}({C_{mb}}^{a}-{{C_{m}}^{a}}_{b})\psi_{a}(0) + \dots
\en
which implies
\be
\partial_{x}\chi_{m}(x)=\lambda^{b}({C_{mb}}^{a}-{{C_{m}}^{a}}_{b})\psi_{a}(x)\, .
\ee
Again, for every broken Chan-Paton symmetry we have a redundant field 
given by the right hand side.

Next we derive the boundary trace anomaly $\theta(x)$, which 
satisfies the conservation equation
\eq
T(x )-\bar T(x )= \partial_{x }\theta(x )
\,.
\en
From (\ref{Tpch}) we calculate
\begin{multline}
\langle [T(x  + i\epsilon)-\bar T(x -i\epsilon)]\psi_{a}(x')\chi_{m}(y)\rangle = 
-2\pi i \lambda^{b}({C_{mb}}^{a}-{{C_{m}}^{a}}_{b})\\
 \Bigl[  \frac{1}{2}\partial^{2}_{x }\delta(x -x') + \partial_{x'}\Bigl(\frac{1}{x'-y}\delta(x-x' )\Bigr) 
 + \frac{1}{(x'-y)^{2}}\delta(x -y)\Bigr]
 \,.
\end{multline}
We can read off the boundary trace anomaly from the highest 
derivative term on the right hand side,
\be
\theta =  -\pi \lambda^{b}\partial_{x }\lambda^{a}(C_{ba}^{m}- 
C_{ab}^{m})\chi_{m} \, . 
\ee
As in the bulk, this term in the trace anomaly gives the anomalous 
dimensions of the redundant fields.

We should note that what we have described is the two dimensional 
physics underlying the so-called string Higgs effect (see e.g. \cite{TdualityReview} for a review).
When the CFT is a string theory compactification,
the spin 1 dimension 1 fields $J_{m}$ give massless gauge fields in space-time.
The charged dimension 2 scalar fields $\phi_{i}$ give massless scalars in 
space-time.
These are the Higgs fields.
A perturbation $\lambda^{i}\phi_{i}$ corresponds to giving a vacuum 
expectation value to the Higgs fields.
The spin 1 fields acquire anomalous dimensions, which correspond to 
the masses of the $W$ bosons.
The redundant fields are the pure gauge directions that are eaten up
by the $W$ bosons.
The boundary case is parallel.

\section{Details of the boundary curvature computation}
\subsection{Proof of identity (\ref{metric_id}) }\label{mproof}
\setcounter{equation}{0}
We need to prove
\eq
\tilde R^{\epsilon}_{abcd} = - \tilde R^{\epsilon}_{abdc}
\en
where
\eq
\tilde R^{\epsilon}_{abcd}
=  \int d\chi \;  F_{\epsilon}(\chi)
\left [  \expvalc{\psi_{b}(0) \psi_{d}(\chi) \psi_{a}(1) \psi_{c}(\infty)}
+ \expvalc{\psi_{b}(1) \psi_{d}(\chi) \psi_{a}(0) \psi_{c}(\infty)}
\right ]
\en
and $F_{\epsilon}(\chi)$ is defined by  (\ref{eq:Fepsilon}).

Making  conformal transformations
\eq
x ' =\left \{
\begin{array}{l@{\qquad}l}
\frac{ (1-\chi)x }{x -\chi} & 
\chi(1-\chi) <0 
\\[1ex]
1-\frac{ (1-\chi)x }{x -\chi}& 
\chi(1-\chi) >0\,,
\end{array}
\right .
\en
we obtain
\eq
\expvalc{\psi_{b}(0) \psi_{d}(\chi) \psi_{a}(1) \psi_{c}(\infty)}
=\left \{
\begin{array}{l@{\qquad}l}
-\expvalc{\psi_{b}(0) \psi_{d}(\infty) \psi_{a}(1) 
\psi_{c}(1-\chi)} & \chi(1-\chi) <0 
\\[1ex]
\expvalc{\psi_{b}(1) \psi_{d}(\infty) \psi_{a}(0) \psi_{c}(\chi)}
& \chi(1-\chi) >0
\end{array}
\right .
\en
\eq
\expvalc{\psi_{b}(1) \psi_{d}(\chi) \psi_{a}(0) \psi_{c}(\infty)}
=\left \{
\begin{array}{l@{\qquad}l}
-\expvalc{\psi_{b}(1) \psi_{d}(\infty) \psi_{a}(0) 
\psi_{c}(1-\chi)} & \chi(1-\chi) <0 
\\[1ex]
\expvalc{\psi_{b}(0) \psi_{d}(\infty) \psi_{a}(1) \psi_{c}(\chi)}
& \chi(1-\chi) >0
\end{array}
\right .
\en
We calculate
\aeq{
\tilde R^{\epsilon}_{abdc}
&= \int\limits_{\chi(1-\chi) <0} d\chi \;  F_\epsilon(1-\chi)\
\left [  \expvalc{\psi_{b}(0) \psi_{d}(\infty) \psi_{a}(1) \psi_{c}(\chi)}
-\expvalc{\psi_{b}(1) \psi_{d}(\infty) \psi_{a}(0) 
\psi_{c}(\chi)}
\right ]
\nonumber\\&\qquad
{}+ \int\limits_{\chi(1-\chi) >0} d\chi \;  F_\epsilon(\chi)
\left [  \expvalc{\psi_{b}(1) \psi_{d}(\infty) \psi_{a}(0) \psi_{c}(\chi)}
- \expvalc{\psi_{b}(0) \psi_{d}(\infty) \psi_{a}(1) \psi_{c}(\chi)}
\right ]
\nonumber \\
&= - \int d\chi \;  F_\epsilon(\chi)
\left [  \expvalc{\psi_{b}(0) \psi_{d}(\infty) \psi_{a}(1) \psi_{c}(\chi)}
- \expvalc{\psi_{b}(1) \psi_{d}(\infty) \psi_{a}(0) 
\psi_{c}(\chi)}
\right ]
\nonumber \\
& = - \tilde R^{\epsilon}_{abcd}
}
which gives (\ref{metric_id}).

\subsection{The functions $F_{\epsilon}(\chi)$ and $\Delta F_{\epsilon}(\chi)$} \label{Fchi}
The function $F_{\epsilon}(\chi)$,
defined by  (\ref{eq:Fepsilon}),
is given in terms of radial integrals
over the region ${\mathcal R_{+}^{\epsilon}}$
at fixed slope $\chi = x_{2}/x_{1}$.
The region ${\mathcal R_{+}^{\epsilon}}$ is defined by (\ref{eq:Repsilonplusregion})
and is depicted in 
figure 1.
\begin{figure*}[h!]
\centering
\includegraphics[width=3.0in]{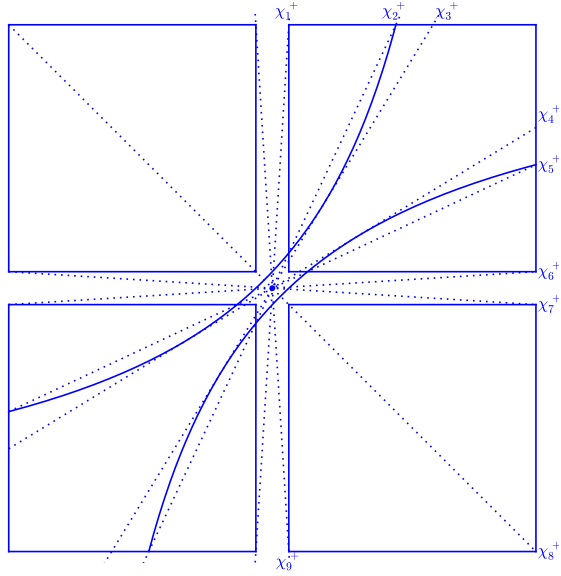}
\caption{The region $\mathcal{R_{+}^{\epsilon}}$}
\end{figure*}
\noindent
The coordinates are $(x_{1},x_{2})$.
The four squares are the regions $\epsilon< |x_{1,2}|<\epsilon^{-1}$.
The upper curve is $x_{2}=y_{+}(x_{1})$ and  the lower curve is 
$x_{2}=y_{-}(x_{1})$ where 
\eq
y_{+}(x) = \frac{x+\epsilon}{1-\epsilon x}\,,
\qquad
y_{-}(x) = \frac{x-\epsilon}{1+\epsilon x} \,,
\qquad
y_{-}(x) = - y_{+}(-x) = y_{+}(x^{-1})^{-1}\,.
\en
The 
region ${\mathcal R_{+}^{\epsilon}}$ consists of the interiors of the 
2 squares on the right, minus the 
portion lying between the curves.  The dotted rays mark the transitions 
where the radial integral over $\mathcal{R_{+}^{\epsilon}}$ is not a smooth function of the 
slope $\chi$.  
The slopes of the dotted rays are labelled $\chi^{+}_{k}$.
The piece-wise continuous function $F_{\epsilon}(\chi)$ is
given in table~\ref{table}.
It is non-smooth at $\chi = 
\chi^{+}_{k}$ and at $\chi = 1- \chi^{+}_{k}$.
The function $\Delta F_{\epsilon}(\eta)$ is defined
in (\ref{eq:DeltaFepsilon}).
\suppressfloats

\renewcommand*\arraystretch{1.24}
{
\begin{table}[ht!]
\begin{tabular}{|l|l|l|}
\multicolumn{1}{c}{$\chi$} & \multicolumn{1}{c}{$F_{\epsilon}(\chi)$}
& \multicolumn{1}{c}{$\Delta F_{\epsilon}(\chi)$}\\[0.5ex]
\hline
\makebox[12em][l]{}
&0&0\\
\hline
\multicolumn{3}{l}{$1-\chi^{+}_{9}=\epsilon^{-2}+1$} \\
\hline
& $\ln(\epsilon^{-2}) -\ln |1-\chi|$ 
&$-\ln (\epsilon^{2}\chi ) = O(\epsilon^{2})$\\
\hline
\multicolumn{3}{l}{$\chi^{+}_{1}=\epsilon^{-2}$}\\
\hline
&$\ln|\chi| -\ln |1-\chi|$& 0\\
\hline
\multicolumn{3}{l}{$\chi^{+}_{2} =\frac2{1-\epsilon^{2}}$} \\
\hline
&
$\ln|\chi| -3\ln |1-\chi|+A \big ( 
\frac{\epsilon^{2}\chi}{(\chi-1)^{2}}\big )$
&$-2\ln |1-\chi|+A\big ( \frac{\epsilon^{2}\chi}{(\chi-1)^{2}}\big )= O(\epsilon^{2})$\\
\hline
\multicolumn{3}{l}{$1-\chi^{+}_{8}=2$}\\
\hline
&$\ln|\chi| -\ln |1-\chi|+A\big ( 
\frac{\epsilon^{2}\chi}{(\chi-1)^{2}}\big )$
&$A \big ( \frac{\epsilon^{2}\chi}{(\chi-1)^{2}}\big )$\\
\hline
\multicolumn{3}{l}{$\chi^{+}_{3} = \left ( \sqrt{1+\epsilon^{2}}+\epsilon \right )^{2}$ }\\
\hline
&$\ln(\epsilon^{-2}) +\ln |1-\chi|$ 
&$-\ln  (\epsilon^{2}\chi )
+ \ln (\chi-1)^{2}
$ \\
\hline
\multicolumn{3}{l}{$1-\chi^{+}_{7}= 1+\epsilon^{2}$}\\
\hline
&0&0\\
\hline
\multicolumn{3}{l}{$1-\chi^{+}_{6} = 1-\epsilon^{2}$}\\
\hline
&$\ln(\epsilon^{-2}) +\ln |1-\chi| $
&$-\ln  (\epsilon^{2}\chi  )
+ \ln (\chi-1)^{2}$\\
\hline
\multicolumn{3}{l}{$\chi^{+}_{4}  = \left ( \sqrt{1+\epsilon^{2}}-\epsilon \right )^{2}$}\\
\hline
&$\ln|\chi| -\ln |1-\chi|+A\big ( 
\frac{\epsilon^{2}\chi}{(\chi-1)^{2}}\big )$&
$A\big ( \frac{\epsilon^{2}\chi}{(\chi-1)^{2}}\big )$\\
\hline
\multicolumn{3}{l}{$1-\chi^{+}_{5} = (1+\epsilon^{2})/2$}\\
\hline
&$3\ln|\chi| -3\ln |1-\chi|$
&$2\ln\big ( \frac{\chi}{1-\chi}\big )+A
\big ( \frac{\epsilon^{2}\chi}{(\chi-1)^{2}}\big )-A\big ( \frac{\epsilon^{2}(1-\chi)}{\chi^{2}}\big )$ 
\\
& $+A\big ( \frac{\epsilon^{2}\chi}{(\chi-1)^{2}}\big )
-A\big ( \frac{\epsilon^{2}(1-\chi)}{\chi^{2}}\big )$
&$= O(\epsilon^{2})$\\
\hline
\multicolumn{3}{l}{$\chi^{+}_{5} = (\chi^{+}_{2})^{-1}= (1-\epsilon^{2})/2$}\\
\hline
&$\ln|\chi| -\ln |1-\chi| -A\big ( 
\frac{\epsilon^{2}(1-\chi)}{\chi^{2}}\big )$
& $-A\big ( \frac{\epsilon^{2}(1-\chi)}{\chi^{2}}\big )$
\\
\hline
\multicolumn{3}{l}{$1-\chi^{+}_{4} = 2\epsilon\sqrt{1+\epsilon^{2}} -2\epsilon^{2} $}\\
\hline
& $\ln(\epsilon^{2}) - \ln |\chi|$
&$\ln (\epsilon^{2}(1-\chi) )
-\ln \chi^{2} $\\
\hline
\multicolumn{3}{l}{$\chi^{+}_{6}=(\chi^{+}_{1})^{-1}= \epsilon^{2}$}\\
\hline
&0&0\\
\hline
\multicolumn{3}{l}{$\chi^{+}_{7}= -\epsilon^{2}$}\\
\hline
&$\ln  (\epsilon^{2} )
-\ln \chi^{2} $
&$
\ln(\epsilon^{2}(1-\chi)) - \ln |\chi|
$\\
\hline
\multicolumn{3}{l}{$1-\chi^{+}_{3} = -2\epsilon\sqrt{1+\epsilon^{2}} 
-2\epsilon^{2}$ }\\
\hline
&$\ln|\chi| -\ln |1-\chi|
-A\big ( \frac{\epsilon^{2}(1-\chi)}{\chi^{2}}\big )$
&$-A\big ( \frac{\epsilon^{2}(1-\chi)}{\chi^{2}}\big )$\\
\hline
\multicolumn{3}{l}{$\chi^{+}_{8} = -1$}\\
\hline
&$3\ln|\chi| -\ln |1-\chi|-A\big ( 
\frac{\epsilon^{2}(1-\chi)}{\chi^{2}}\big )$
&$2\ln|\chi| -A\big ( \frac{\epsilon^{2}(1-\chi)}{\chi^{2}}\big )= O(\epsilon^{2})$\\
\hline
\multicolumn{3}{l}{$1- \chi^{+}_{2} =  ( 1+\epsilon^{2})/(\epsilon^{2}-1) $}
\\
\hline
&$\ln|\chi| -\ln |1-\chi|$ &0\\
\hline
\multicolumn{3}{l}{$1-\chi^{+}_{1} =-\epsilon^{-2}+1$}\\
\hline
&$\ln|\chi|+\ln (\epsilon^{2})$
&$\ln|1-\chi|+\ln (\epsilon^{2})= O(\epsilon^{2})$\\
\hline
\multicolumn{3}{l}{$\chi^{+}_{9}=(\chi^{+}_{7})^{-1} = -\epsilon^{-2}$}\\
\hline
 & 0&0\\
\hline
\end{tabular}
\caption{The functions $F_{\epsilon}(\chi)$ and $\Delta F_{\epsilon}(\chi)$ }
\label{table}
\end{table}
}
\renewcommand*\arraystretch{1}
\newpage
In table 1, the function $A(s)$ is
\eq
A(s) = - 2 \ln \left (\frac12 +\sqrt{\frac14 -s} \right )
= 2s + s^{2} + O(s^{3})
\,.
\en
The left column of the table lists the values of $\chi$ where $F_{\epsilon}(\chi)$ 
is non-smooth,  arranged in decreasing order from $\chi=+\infty$ to 
$\chi = -\infty$.
The rows between two adjacent thresholds give the values of 
$F_{\epsilon}(\chi)$ and $\Delta F_{\epsilon}(\chi)$ 
for $\chi$ in the corresponding interval.
\newpage

\subsection{Computation of $E_{abcd}$}\label{Edetails}
The error term $E_{abcd}$ is defined by (\ref{Edef}) as an integral 
over $\chi$ of four-point functions
weighted by $\Delta F_{\epsilon}(\chi)$, which is given 
in table 1.
We need to derive equation (\ref{eq:Eerrorterm})
which gives the asymptotic behavior of $E_{abcd}$  in the limit $\epsilon\rightarrow 0$.

First we note the reflection symmetry
\be 
\Delta F_{\epsilon}(1-\chi) = - \Delta 
F_{\epsilon}(\chi)\,.
\ee 
Next we note that the $\chi$-intervals where $\Delta F_{\epsilon}$ is identically zero 
of course make no contribution.
By inspection, we see that the $\chi$-intervals where $\Delta 
F_{\epsilon}$ is explicitly written as ${\cal O}(\epsilon^{2})$ in 
the table
can also be neglected,
because the four point functions are bounded there.
We are left with four  $\chi$-intervals lying in the region
$-1 \le \chi \le (1-\epsilon^{2})/2$
and four  $\chi$-intervals lying in the reflected region.
Now we can use the reflection symmetry to write
\be
E_{abcd}=\int\limits_{-1}^{(1-\epsilon^{2})/2}d\chi \Delta F_{\epsilon}(\chi)[G_{abcd}(\chi)-G_{abcd}(1-\chi)]
\ee
where 
\be
G_{abcd}(\chi)=\langle \psi_{b}(0)\psi_{c}(\chi)\psi_{a}(1)\psi_{d}(\infty)\rangle_{c} 
- \langle \psi_{a}(0)\psi_{c}(\chi)\psi_{b}(1)\psi_{d}(\infty)\rangle_{c} \, .  
\ee
In this region, $\Delta F_{\epsilon} \rightarrow 0$ except for a 
shrinking neighborhood of $\chi=0$ where it diverges only logarithmically.
Therefore
non-negligible contributions to $E_{abcd}$  come  only
from the singularities ni the four-point functions associated with 
relevant operators runing in the intermediate channels.
Thus, up to terms vanishing in the limit $\epsilon\to 0$  we have 
\be 
E_{abcd}= \sum_{a'}[G_{a'}^{+}E^{+}(\Delta_{a'}) + G_{a'}^{-}E^{-}(\Delta_{a'})]
\ee
where 
\aeq{
E^{+}(\Delta) &= \int_{\epsilon^{2}}^{\frac{1-\epsilon^{2}}{2}}
d\chi \; \Delta F_\epsilon(\chi)\chi^{\Delta-2}\,,
\\
E^{-}(\Delta) &= \int_{-1}^{-\epsilon^{2}}
d\chi \; \Delta F_\epsilon(\chi)(-\chi)^{\Delta-2}
=
\int_{\epsilon^{2}}^{1}
d\chi \; \Delta F_\epsilon(-\chi)\chi^{\Delta-2}\,,
}
\aeq{
 G_{a'}^{+} &= [C^{a'}_{cb} C_{daa'} - C^{a'}_{ac} C_{bda'} ] 
(1-a\leftrightarrow b)\,,
\\[1ex]
 G_{a'}^{-} &= [C^{a'}_{bc} C_{daa'} - C^{a'}_{ca} C_{bda'} ]
(1-a\leftrightarrow b)\,.
}
Changing the integration variable to
$\chi = \epsilon u$ we calculate
\aeq{
E^{+}(\Delta) &= \epsilon^{\Delta-1}\int_{\epsilon}^{\frac{1-\epsilon^{2}}{2\epsilon}}
du \; k_{+}(u) u^{\Delta-2}
= \epsilon^{\Delta-1}\int_{\epsilon}^{\frac{1-\epsilon^{2}}{2\epsilon}}
du \; k_{+}(u) \partial_{u} \frac{u^{\Delta-1}}{\Delta-1}
\\[1ex]
E^{-}(\Delta) &= 
\epsilon^{\Delta-1}\int_{\epsilon}^{\frac1\epsilon}
du \; k_{-}(u)u^{\Delta-2}
=
\epsilon^{\Delta-1}\int_{\epsilon}^{\frac1\epsilon}
du \; k_{-}(u)\partial_{u} \frac{u^{\Delta-1}}{\Delta-1}
}
where
\aeq{
k_{+}(u) &= \Delta F_\epsilon(\epsilon u)
= \left \{
\begin{array}{l@{\qquad}l}
\ln \left (\frac{1-\epsilon u}{u^{2}} \right ) & \epsilon \le u \le  
2\sqrt{1+\epsilon^{2}} -2\epsilon \\[1ex]
-A(\frac{1-\epsilon u}{u^{2}}) &  2\sqrt{1+\epsilon^{2}} -2\epsilon 
\le u \le \frac{1-\epsilon^{2}}{2\epsilon}
\end{array}
\right .
\\[2ex]
k_{-}(u) &= \Delta F_\epsilon(-\epsilon u)
= \left \{
\begin{array}{l@{\qquad}l}
\ln \left (\frac{1+\epsilon u}{u^{2}} \right ) & \epsilon \le u \le  
2\sqrt{1+\epsilon^{2}} +2\epsilon \\[1ex]
-A(\frac{1+\epsilon u}{u^{2}}) & 2\sqrt{1+\epsilon^{2}} +2\epsilon\le 
u \le \frac1\epsilon\,.
\end{array}
\right .
}
Integrating by parts, using the fact that the $k_{\pm}(u)$ are continuous 
within the range of integration, and dropping the contributions from 
the upper boundaries because they 
are $O(\epsilon^{2})$, we get
\aeq{
E^{+}(\Delta) &= \epsilon^{\Delta-1}
\left [ 
k_{+}(\epsilon)\frac{\epsilon^{\Delta-1}}{1-\Delta}
+ \int_{\epsilon}^{\frac{1-\epsilon^{2}}{2\epsilon}}
du \;  k_{+}'(u)\frac{u^{\Delta-1}}{1-\Delta}
\right ] 
\\[1ex]
E^{-}(\Delta) &= 
\epsilon^{\Delta-1}
\left [ 
k_{-}(\epsilon)\frac{\epsilon^{\Delta-1}}{1-\Delta}
+ \int_{\epsilon}^{\frac1\epsilon}
du \; k_{-}'(u)\frac{u^{\Delta-1}}{1-\Delta}
\right ] 
\,.
}
Since
\eq
k_{\pm}(\epsilon) =  \ln (\epsilon^{-2}) + O(\epsilon^{2})\,,
\en
we can write
\eq
E^{\pm}(\Delta) = \ln(\epsilon^{-2})\frac{(\epsilon^{2})^{\Delta-1}}{1-\Delta}
+ O(\epsilon^{2\Delta}) +I^{\pm}(\Delta)
\en
with 
\aeq{
I^{+}(\Delta) &= \epsilon^{\Delta-1}
\int_{\epsilon}^{\frac{1-\epsilon^{2}}{2\epsilon}}
du \;  k_{+}'(u)\frac{u^{\Delta-1}}{1-\Delta}
\\[1ex]
I^{-}(\Delta) &= 
\epsilon^{\Delta-1}
\int_{\epsilon}^{\frac1\epsilon}
du \; k_{-}'(u)\frac{u^{\Delta-1}}{1-\Delta}\,.
}
Analyzing the behaviour of these integrals in the limit $\epsilon \to 
0$ we find,
up to terms vanishing in the limit $\epsilon \to 0$,
\aeq{
E^{+}(\Delta)
=
E^{-}(\Delta)
&=
\ln (\epsilon^{-2})\frac{(\epsilon^{2})^{\Delta-1}}{1-\Delta} 
-2 \frac{(\epsilon^{2})^{\Delta-1}}{(1-\Delta)^{2}}
\nonumber\\[1ex]
&\quad {}+
\frac{\epsilon^{\Delta-1}}{1-\Delta}
\int_{2}^{\infty}
du \; \left [
-\partial_{u} A(u^{-2}) + 2 u^{-1}
\right ]u^{\Delta-1}
\,.
}
Therefore
\eq
E_{abcd}=
\sum_{a'}
E^{+}(\Delta_{a'}) \left ( G^{+}_{a'} 
+ G^{-}_{a'}
\right )
=
\sum_{a'}
E^{+}(\Delta_{a'}) \left (
C^{a'}_{(ad)}C_{(bc)a'}-C^{a'}_{(ac)}C_{(bd)a'}
\right )
\en
which is equation (\ref{eq:Eerrorterm}).

\section{Distributional correlators of currents} \label{app1}
\setcounter{equation}{0}
In this appendix we construct the
distributional three-point and four-point correlation functions of currents for the
D0 brane example discussed in section~\ref{sec_D0}. 
By translation invariance, the three-point function is a distribution in two 
real variables and the four-point function is a distribution in three
real variables.
Along the way we derive some useful identities on distributions.

\subsection{Distributions in two variables and the three-point function}

We define a 
distribution ${\rm PV}\frac1{xy}$ in the two real variables $x$ and $y$
by its action on test functions $f (x,y)$,
\be
\big (f , {\mathrm {PV}}\frac{1}{xy}\big ) = \lim_{\epsilon \to 0} \iint\limits_{|x|, |y|\ge \epsilon} 
\frac{f (x,y)}{xy}
\ee
which is equivalent to
\eq
(f , {\rm PV}\frac{1}{xy}) 
= \int\limits_{0}^{\infty}dx \int\limits_{0}^{\infty}dy\, 
\frac{1}{xy}(1-R_{x})(1-R_{y})f (x,y)
\en
where 
\bea
&&R_{x}: (x,y) \mapsto (-x,y) \\
&& R_{y}:(x,y) \mapsto (x,-y)\,.
\eea
Next  define 
\be 
(f , {\rm PV}\frac{1}{x(y-x)}) = \lim_{\epsilon \to 0} \iint\limits_{|x|, |y-x|\ge \epsilon} 
\frac{f (x,y)}{x(y-x)}
\ee
\be 
(f , {\rm PV}\frac{1}{y(x-y)}) = \lim_{\epsilon \to 0} \iint\limits_{|y|, |y-x|\ge \epsilon} 
\frac{f (x,y)}{y(x-y)}
\ee
which are equivalent to
\be 
(f , {\rm PV}\frac{1}{x(y-x)}) = \int\limits_{0}^{\infty}dx \int\limits_{0}^{\infty}dy\, 
\frac{1}{xy}(1-R_{x})(1-R_{y}) f (x,y+x)  \, , 
\ee
\be 
(f , {\rm PV}\frac{1}{y(x-y)}) = \int\limits_{0}^{\infty}dx \int\limits_{0}^{\infty}dy\, 
\frac{1}{xy}(1-R_{x})(1-R_{y}) f (x+y,y) \, .  
\ee
The following useful identities follow directly from the definitions
\be \label{id1}
{\rm PV}\frac{1}{xy} - {\rm PV}\frac{1}{x(y-x)} - {\rm PV}\frac{1}{y(x-y)} 
= \pi^{2}\delta(x)\delta(y)\, , 
\ee
\be \label{id2}
{\rm PV}\frac{1}{xy} - {\rm PV}\frac{1}{x(y+x)} - {\rm PV}\frac{1}{y(x+y)} 
= -\pi^{2}\delta(x)\delta(y)\, . 
\ee

We now turn to constructing the distributional three-point function of currents on the boundary.
At finite separations the three-point function on the boundary is
\be 
\langle J^{a}(x _{1})J^{b}(x _{2})J^{c}(x _{3})\rangle = - i k  f^{abc} 
\frac{1}{x _{12}x _{13}x _{32}} \,.
\ee
We want a distributional regularization
\be 
\Bigl[\frac{1}{x _{12}x _{13}x _{32}}   \Bigr]
\ee
of the rational function,
which must be fully antisymmetric in $x _{1}$, $x _{2}$, $x _{3}$
because the three-point function is symmetric.
By translation invariance, we 
can think of such a distribution as a distribution in two variables 
$x _{2}$ and $x _{3}$,
treating $x _{1}$ as a parameter. 

Define
\be
\Bigl[ \frac{1}{x _{12}x _{13}x _{32}} \Bigr]_{1}  = \partial_{2} {\rm PV} \frac{1}{x _{32}x _{13}}  
+ \partial_{3} {\rm PV} \frac{1}{x _{32}x _{12}}  
\,.
\ee
  Using the identities (\ref{id1}), (\ref{id2}) we find that this distribution transforms 
  in the following way under permutations $\sigma_{12}$, $\sigma_{23}$:
\aeq{
\sigma_{12}\Bigl[ \frac{1}{x _{12}x _{13}x _{32}} \Bigr]_{1} &=
- \Bigl[ \frac{1}{x _{12}x _{13}x _{32}} \Bigr]_{1} + \pi^{2}\partial_{3}(\delta(x _{12})\delta(x _{13}))
\\[1ex]
\sigma_{23}\Bigl[ \frac{1}{x _{12}x _{13}x _{32}} \Bigr]_{1} &=
- \Bigl[ \frac{1}{x _{12}x _{13}x _{32}} \Bigr]_{1} 
\,.
}
It follows that the distribution 
\be
\label{distr_3}
\Bigl[\frac{1}{x _{12}x _{13}x _{32}}   \Bigr]= \partial_{2} {\rm PV} \frac{1}{x _{32}x _{13}}  
+ \partial_{3} {\rm PV} \frac{1}{x _{32}x _{12}} + \frac{\pi^{2}}{3}(\partial_{2}-\partial_{3})
\delta(x _{12})\delta(x _{13}) 
\ee
is fully antisymmetric. We thus set
\be \label{3pt2}
\langle J^{a}(x _{1})J^{b}(x _{2})J^{c}(x _{3})\rangle = -i k  f^{abc} 
\Bigl[ \frac{1}{x _{12}x _{13}x _{32}} \Bigr]\,.
\ee

\subsection{Distributions in three variables and the four-point function}
We define the following distributions in three real variables $x,y,z$:
\be 
\big (f , {\rm PV}\frac{1}{xyz}\big ) = 
\int\limits_{0}^{\infty}dx \int\limits_{0}^{\infty}dy \int\limits_{0}^{\infty}dz \, 
\frac{1}{xyz}(1-R_{x})(1-R_{y})(1-R_{z})f (x,y,z) \, , 
\ee
\begin{multline}
\big ( f , {\rm PV}\frac{1}{(x-x')(x-y)(y-z)}\big ) =
\int\limits_{0}^{\infty}\!\!dx \int\limits_{0}^{\infty}\!\!dy \int\limits_{0}^{\infty}\!\!dz \, 
\frac{1}{xyz}
\\[1ex]
 (1-R_{x})(1-R_{y})(1-R_{z})f (x'-x,x'-x-y,x'-x-y-z)
\end{multline}
where
\be
R_{z}:(x,y,z)\mapsto (x,y,-z) \, 
\ee
and $x'$ is a parameter.
Distributions with other factors in the denominator are defined 
analogously.
Identities of the following type hold
\begin{multline}
{\rm PV}\frac{1}{(x-x')(x-y)(y-z)} = {\rm PV}\frac{1}{(x-x')(x-y)(x-z)} \\
{} +  {\rm PV}\frac{1}{(x-x')(x-z)(y-z)} - \pi^{2}\delta(x-y)\delta(y-z){\rm PV}\frac{1}{x'-x}
\,.
\end{multline}
That is, the identities (\ref{id1}) and (\ref{id2}) can be used inside the PV symbol. 

The connected four-point function of currents on the boundary is, at 
separated points,
\eq
 \label{4ptc}
 \langle J^{a}(x _1)J^{b}(x _2)J^{c}(x _3)J^{d}(x _4)\rangle_{c} 
 = \frac{k{f^{ac}}_{s}f^{bsd}}{x _{13}x _{12}x _{34}x _{24}} 
+ \frac{k{f^{ad}}_{s}f^{bcs}}{x _{14}x _{12}x _{23}x _{43}}  \, .
\en
We need to extend the rational functions to distributions in three 
variables, which we take to be  $x _{2}, x _{3}, x _{4}$, leaving 
$x_{1}$ as a parameter.
The full distributional four-point function must be symmetric under simultaneous permutations of 
$x _{i}$ and the corresponding group indices.
By a slight abuse of terminology we will refer to this symmetry 
as crossing symmetry.

We start by defining 
\bea 
&& D_{1}\equiv \Bigl[\frac{1}{x _{12}x _{23}x _{34}x _{42}}  \Bigr]_{1}
= - \partial_{4}{\rm PV}\frac{1}{x _{12}x _{34}x _{23}} - 
\partial_{3}{\rm PV}\frac{1}{x _{12}x _{42}x _{43}} \, ,\\[1ex]
&& D_{2}\equiv \Bigl[\frac{1}{x _{13}x _{23}x _{34}x _{42}}  \Bigr]_{1}
= - \partial_{4}{\rm PV}\frac{1}{x _{13}x _{24}x _{23}} - \partial_{2}{\rm PV}\frac{1}{x _{13}x _{42}x _{34}}\, , \\[1ex]
&& D_{3}\equiv \Bigl[\frac{1}{x _{14}x _{23}x _{34}x _{42}}  \Bigr]_{1}
= - \partial_{3}{\rm PV}\frac{1}{x _{14}x _{23}x _{42}} - \partial_{2}{\rm PV}\frac{1}{x _{14}x _{32}x _{34}}\, ,
\eea
and then defining
\bea\label{distrs}
&& \Bigl[ \frac{1}{x _{13}x _{12}x _{34}x _{24}}  \Bigr]_{1} = D_{2}-D_{1}\, , \\[1ex]
&& \Bigl[\frac{1}{x _{14}x _{12}x _{23}x _{34}} \Bigr]_{1} = D_{3} - D_{1}\, , \\[1ex]
&& \Bigl[ \frac{1}{x _{13}x _{14}x _{23}x _{24}}  \Bigr]_{1} = D_{3}-D_{2} \, . 
\eea
The distributions $D_2 - D_1 $ and $D_3-D_1$ regularize the rational functions 
which appear in (\ref{4ptc}). We next turn to their behaviour 
under permutations. We find 
\aeq{
\sigma_{12}(D_{2}-D_{1}) & = D_{1}-D_{3} + \Xi_{1}\, , \\[1ex]
\sigma_{13}(D_{2}-D_{1})&= D_{3} - D_{2} + \Xi_{2}\, , \\[1ex]
\sigma_{14}(D_{3}-D_{1}) &= D_{2}-D_{3} + \Xi_{3}\, ,
}
where $\Xi_{i}$ are contact terms which can be computed using 
(\ref{id1}) and (\ref{id2}):
\aeq{
 \Xi_{1}&= -\pi^{2} \partial_{4}\Bigl( \delta(x _{14})\delta(x _{34}){\rm PV}\frac{1}{x _{23}} 
+  \delta(x _{13})\delta(x _{23}){\rm PV}\frac{1}{x _{34}}  \Bigr) \nonumber \\
&\quad{} + \pi^{2} \partial_{3}\Bigl( \delta(x _{21})\delta(x _{41}){\rm PV}\frac{1}{x _{43}}  
- \delta(x _{23})\delta(x _{34}){\rm PV}\frac{1}{x _{41}} \Bigr) \, ,
\nonumber\\
\Xi_{2}&= -\pi^{2}\partial_{4}
\Bigl( \delta(x _{42})\delta(x _{41}){\rm PV}\frac{1}{x _{32}} + \delta(x _{31})\delta(x _{21}){\rm PV}\frac{1}{x _{24}}   
\Bigr) \nonumber \\
& \quad{}+ \pi^{2}\partial_{2}\Bigl( \delta(x _{31})\delta(x _{14}){\rm PV}\frac{1}{x _{42}} 
- \delta(x _{32})\delta(x _{42}){\rm PV}\frac{1}{x _{41}} \Bigr)\, ,
\nonumber\\
\Xi_{3}&= \pi^{2}\partial_{3}\Bigl(\delta(x _{31})\delta(x _{23}){\rm PV}\frac{1}{x _{42}}
+ \delta(x _{41})\delta(x _{12}){\rm PV}\frac{1}{x _{23}}
 \Bigr)  \nonumber \\
& \quad{}+ \pi^{2}\partial_{2}\Bigl(-\delta(x _{41})\delta(x _{31}){\rm PV}\frac{1}{x _{32}}
+ \delta(x _{42})\delta(x _{23}){\rm PV}\frac{1}{x _{31}}
 \Bigr)\,.
}
To satisfy the crossing symmetry we modify $D_{2} - D_{1}$ according to the ansatz
\eq
\Bigl[\frac{1}{x _{13}x _{12}x _{34}x _{24}}  \Bigr]=D_{2} - D_{1} + \xi
\en
with
\begin{multline}
\xi = A_{1}\delta_{234}\Bigl[\frac{1}{x _{12}^{2}}  \Bigr] 
+ A_{2} \delta_{134} \Bigl[\frac{1}{x _{12}^{2}}  \Bigr]
+ A_{3} \delta_{124} \Bigl[\frac{1}{x _{13}^{2}}  \Bigr]
+ A_{4} \delta_{123} \Bigl[\frac{1}{x _{41}^{2}}  \Bigr] + B_{123}\partial_{3}(\delta_{123}{\rm PV}\frac{1}{x _{43}}) \\
+ B_{132}\partial_{2}(\delta_{123}{\rm PV}\frac{1}{x _{43}})
+ B_{124}\partial_{4}(\delta_{124}{\rm PV}\frac{1}{x _{34}}) 
+ B_{142}\partial_{2}(\delta_{124}{\rm PV}\frac{1}{x _{34}})
+ B_{134}\partial_{4}(\delta_{134}{\rm PV}\frac{1}{x _{24}}) \\
+ B_{143}\partial_{3}(\delta_{134}{\rm PV}\frac{1}{x _{34}})
+ B_{234}\partial_{4}(\delta_{234}{\rm PV}\frac{1}{x _{14}}) 
+ B_{243}\partial_{3}(\delta_{234}{\rm PV}\frac{1}{x _{13}})  
\,,
\end{multline}
defining
\eq
\delta_{ijk}=\delta(x _{ij})\delta(x _{jk})\, , \qquad 
\Bigl[\frac{1}{x _{ij}^{2}}\Bigr] = \partial_{j}{\rm PV}\frac{1}{x _{ij}} \, .
\en
Similarly we make the ansatz
\eq
\Bigl[\frac{1}{x _{14}x _{12}x _{23}x _{34}} \Bigr]
= D_{3} - D_{1} - \eta
\en
where $\eta$ has the same form as $\xi$ above with the coefficients $A_{i}$ replaced by $P_{i}$ and 
 $B_{ijk}$ replaced by $Q_{ijk}$. The crossing symmetry requirement implies the equations 
 \bea 
 && \eta - \sigma_{12}\xi = \Xi_{1} \,,\\
 && \eta + \xi + \sigma_{13}\xi = - \Xi_{2}\,,\\
 && \xi + \eta + \sigma_{14}\eta = \Xi_{3}\,.
 \eea 
   Solving these equations we obtain that the only nonvanishing coefficients present in 
   $\xi$ and $\eta$ are 
   \bea
   && A_{1}=A_{2}=A_{3}= - \frac{\pi^{2}}{3}\, ,\quad A_{4} = \frac{2\pi^{2}}{3} \, , \\
   && P_{1}=P_{2}=P_{4}=-\frac{\pi^{2}}{3}\, , \quad P_{3} = \frac{2\pi^{2}}{3}\, , \\
   && B_{234}=Q_{243} = \pi^{2} \,.
   \eea
   The full distributional four-point function is
\eq \label{4pt_distr}
\langle J^{a}(t_1)J^{b}(t_2)J^{c}(t_3)J^{d}(t_4)\rangle_{c} 
= k{f^{ac}}_{s}f^{bsd}\Bigl[\frac{1}{x _{13}x _{12}x _{34}x _{24}}\Bigr]
+ k{f^{ad}}_{s}f^{bcs} \Bigl[\frac{1}{x _{14}x _{12}x _{23}x _{43}}\Bigr]
\en
where
\aeq{
\Bigl[\frac{1}{x _{13}x _{12}x _{34}x _{24}}\Bigr] &=
 \Bigl[\frac{1}{x _{13}x _{12}x _{34}x _{24}}\Bigr]_{1}
    + \pi^{2}\partial_{4}(\delta_{234}){\rm PV}\frac{1}{x _{12}}
\nonumber \\[1ex]
& \quad {}  +  \frac{\pi^{2}}{3}\left ( 2\delta_{123}\Bigl[ \frac{1}{x _{14}^{2}} \Bigr] 
   -\delta_{234}\Bigl[ \frac{1}{x _{12}^{2}} \Bigr] -\delta_{134}\Bigl[ \frac{1}{x _{32}^{2}} \Bigr]
   -\delta_{124}\Bigl[ \frac{1}{x _{32}^{2}} \Bigr] \right )\,,
\\[1ex]
 \Bigl[\frac{1}{x _{14}x _{12}x _{23}x _{43}}\Bigr] &=
  \Bigl[\frac{1}{x _{14}x _{12}x _{23}x _{43}}\Bigr]_{1} 
  -\pi^{2}\partial_{3}\delta_{234}{\rm PV}\frac{1}{x _{12}} 
\nonumber \\[1ex]
& \quad {}
+  \frac{\pi^{2}}{3}\left ( -2\delta_{124}\Bigl[ \frac{1}{x _{13}^{2}} \Bigr] 
   +\delta_{234}\Bigl[ \frac{1}{x _{12}^{2}} \Bigr] +\delta_{134}\Bigl[ \frac{1}{x _{32}^{2}} \Bigr]
   +\delta_{123}\Bigl[ \frac{1}{x _{41}^{2}} \Bigr] \right )\,.
}


\end{document}